\documentclass[11pt]{jdg-p}
\usepackage{amsmath,amssymb}%

\usepackage{graphicx}  

\usepackage{color}

\usepackage{psfrag}
\usepackage{epsfig}
\usepackage{a4}
\usepackage{amssymb,latexsym}
\usepackage[mathscr]{eucal}
\usepackage{cite}
\usepackage{url}

\DeclareFontFamily{OT1}{rsfs}{} \DeclareFontShape{OT1}{rsfs}{m}{n}{
<-7> rsfs5 <7-10> rsfs7 <10-> rsfs10}{}
\DeclareMathAlphabet{\mycal}{OT1}{rsfs}{m}{n}
\def\scri{{\mycal I}}%
\newcommand{\mychitwo}{\mathring \chi }
\newcommand{\mychithree}{\tilde\chi }
\newcommand{\mychione}{\chi }
\newcommand{\myalph}{a}
\newcommand{\mygamma}{\sigma}
\newcommand{\mutheta}{\theta}
\newcommand{\myrho}{\theta}
\newcommand{\dmyrho}{\dot\theta}

\newcommand{\tmcM}{\,\,\,\,\widetilde{\!\!\!\!\mcM}}%
\newcommand{\tg}{\widetilde{g}}%
\newcommand{\obstruction}{--compatibility}%

\newcommand{\kk}[1]{}

\newcommand{\eean}{\nonumber\end{eqnarray}}

\newcommand{\mcMext}{\Mext}

\def\K0{\phi^{K_0}}

\def\X.{\phi^{X}\cdot}

\newcommand{\hmcM}{\,\,\,\widehat{\!\!\!\mcM}}

{\catcode `\@=11 \global\let\AddToReset=\@addtoreset}
\AddToReset{equation}{section}
\renewcommand{\theequation}{\thesection.\arabic{equation}}
{\catcode `\@=11 \global\let\AddToReset=\@addtoreset}
\AddToReset{figure}{section}
\renewcommand{\thefigure}{\thesection.\arabic{figure}}

\newcommand{\nopcite}[1]{}

  {\renewcommand{\theenumi}{\roman{enumi}}
   \renewcommand{\labelenumi}{(\theenumi)}}

\newcommand{\mcE}{{\mycal E}}

\newcommand{\nablash}{\nabla{\kern -.75 em
     \raise 1.5 true pt\hbox{{\bf/}}}\kern +.1 em}
\newcommand{\Deltash}{\Delta{\kern -.69 em
     \raise .2 true pt\hbox{{\bf/}}}\kern +.1 em}
\newcommand{\Rslash}{R{\kern -.60 em
     \raise 1.5 true pt\hbox{{\bf/}}}\kern +.1 em}

\newcommand{\mcO}{{\mycal O}}

\newcommand{\hyp}{{\mycal S}}

\newcommand{\mcM}{{\mycal M}}

\newcommand{\mcH}{{\mycal H}}

\newcommand{\bea}{\begin{eqnarray}}
\newcommand{\beaa}{\begin{eqnarray*}}
\newcommand{\bean}{\begin{eqnarray}\nonumber}

\newcommand{\bel}[1]{\begin{equation}\label{#1}}
\newcommand{\beal}[1]{\begin{eqnarray}\label{#1}}
\newcommand{\beadl}[1]{\begin{deqarr}\label{#1}}
\newcommand{\eeadl}[1]{\arrlabel{#1}\end{deqarr}}
\newcommand{\eeal}[1]{\label{#1}\end{eqnarray}}
\newcommand{\eead}[1]{\end{deqarr}}
\newcommand{\eea}{\end{eqnarray}}
\newcommand{\eeaa}{\end{eqnarray*}}

\newcommand{\be}{\begin{equation}}
\newcommand{\ee}{\end{equation}}

\newcommand{\eq}[1]{\eqref{#1}}
\newcommand{\Eq}[1]{Equation~(\ref{#1})}

\newtheorem{defi}{\sc Coco\rm}[section]

\newtheorem{Theorem}[defi]{\sc Theorem\rm}

\newtheorem{Proposition}[defi]{\sc Proposition\rm}

\newtheorem{Lemma}[defi]{\sc Lemma\!\rm}

\theoremstyle{remark}

\def \R {\Reel}

\def \Nat{\mathbb{N}}
\def \Z{\mathbb{Z}}
\def \N {\Nat}

\newcounter{mnotecount}[section]

\renewcommand{\themnotecount}{\thesection.\arabic{mnotecount}}

\newcommand{\mnote}[1]
{\protect{\stepcounter{mnotecount}}$^{\mbox{\footnotesize $%
\!\!\!\!\!\!\,\bullet$\themnotecount}}$ \marginpar{
\raggedright\tiny\em $\!\!\!\!\!\!\,\bullet$\themnotecount: #1} }

\newcommand{\ednote}[1]{}

\definecolor{bluem}{rgb}{0,0,0.5}

\definecolor{mycolor}{cmyk}{0.5,0.1,0.5,0}
\definecolor{michel}{rgb}{0.5,0.9,0.9}

\definecolor{turquoise}{rgb}{0.25,0.8,0.7}
\definecolor{bluem}{rgb}{0,0,0.5}

\definecolor{MDB}{rgb}{0,0.08,0.45}
\definecolor{MyDarkBlue}{rgb}{0,0.08,0.45}

\definecolor{MLM}{cmyk}{0.1,0.8,0,0.1}
\definecolor{MyLightMagenta}{cmyk}{0.1,0.8,0,0.1}

\definecolor{HP}{rgb}{1,0.09,0.58}

\newcommand{\Mext}{\mcM_{\mathrm{ext}}}

\newcommand{\doc}{\langle\langle \mcMext\rangle\rangle}

\def\exp{\,\mbox{exp}}

\def\emph#1{{\it #1}}
\def\textbf#1{{\bf #1}}

\def\AA{{\mycal  A}}

\def\L{{\bf L}}
\def\O{{\bf O}}

\def\R{{\mathbb R}}

\def\K{{\bf K}}

\def\2{{\overline 2}}

\newcommand{\beqa}{\begin{eqnarray}}
\newcommand{\eeqa}{\end{eqnarray}}

\newcommand{\oldxifour}{\xi_3{}}%
\newcommand{\oldxitwo}{\xi_1{}}%
    \newcommand{\mydot}[1]{\frac{d#1}{ds}}%
     \newcommand{\bmydot}[2]{\left(\frac{d#1}{ds}\right)#2}%
     \newcommand{\mhatg}{g}

\newcommand{\thyp}{\,\,\,\widetilde{\!\!\! \hyp}}

\newcommand{\mydotz}[1]{\frac{d #1}{dz}}%

\begin{document}
\title{Maximal analytic extensions of the Emparan-Reall black ring}

\author{Piotr T.~Chru\'sciel}
\address{LMPT, F\'ed\'eration Denis Poisson, Tours \\ Mathematical
Institute and Hertford College, Oxford}
\author{Julien Cortier}
\address{Institut de Math\'ematiques et de
Mod\'elisation de Montpellier\\ Universit\'e Montpellier 2}%
\thanks{The authors are
grateful to the Mittag-Leffler Institute, Djursholm, Sweden,
for hospitality and financial support during a significant part
of work on this paper.}
\begin{abstract}
We construct a Kruskal-Szekeres-type analytic extension of the
Emparan-Reall black ring, and investigate its geometry. We
prove that the extension is maximal, globally hyperbolic, and
unique within a natural class of extensions. The key to those
results is the proof that causal geodesics are either complete,
or approach a singular boundary in finite affine time.
Alternative maximal analytic extensions are also constructed.
\end{abstract}
\maketitle{}
\date{}

\tableofcontents

\section{Introduction}
 \label{SI}

The Emparan-Reall~\cite{EmparanReall} metrics form a remarkable
class of vacuum black hole solutions of Einstein equations in
dimension $4+1$. Some aspects of their global properties have
been studied in~\cite{EmparanReall}, where it was shown that
the solution contains a \emph{Killing horizon} with $S^2\times
S^1 \times \R$ topology. The aim of this work is to point out
that the \emph{event horizon} coincides with the Killing
horizon, and therefore also has this topology; and to construct
an analytic extension with a bifurcate Killing horizon; and to
establish some global properties of the extended space-time.
The extension resembles closely the Kruskal-Szekeres extension
of the Schwarzschild space-time, with a bifurcate Killing
horizon, a black hole singularity, a white hole singularity,
and two asymptotically flat regions. We show that causal
geodesics in the extended space-time are either complete or
reach a singularity in finite time. This implies maximality of
our extension. We also show global hyperbolicity,  present
families of alternative extensions, establish uniqueness of our
extension within a natural class, and verify existence of a
conformal completion at null infinity.

\section{The Emparan-Reall space-time}
 \label{SERm}

In local coordinates the Emparan-Reall metric can be written in
the form
\begin{eqnarray}
   \label{ERmetric}
   \label{metric20}\phantom{xxxxxx}
g
 &=&-\frac{F(x)}{F(z)}\left(dt+\sqrt{\frac{\nu}{\xi_{F}}}\frac{\xi_{1}-z}{A}d\psi\right)^{2}
   +\frac{F(z)}{A^{2}(x-z)^{2}}\times
   \\
 &&
   \left[-F(x)\left(\frac{dz^{2}}{G(z)}+\frac{G(z)}{F(z)}d\psi^{2}\right)\right.
   \left.+F(z)\left(\frac{dx^{2}}{G(x)}+\frac{G(x)}{F(x)}d\varphi^{2}\right)\right]\;
   ,\nonumber
\end{eqnarray}
   where $A>0$, $\nu$ et $\xi_{F}$ are constants, and
\begin{eqnarray}F(\xi)=1-\frac{\xi}{\xi_{F}}\;, \\
                            G(\xi)=\nu \xi^{3}-\xi^{2}+1=\nu
                            (\xi-\xi_{1})(\xi-\xi_{2})(\xi-\xi_{3})\;,
            \end{eqnarray}
are polynomials, with $\nu$  chosen so that
$\xi_{1}<0<\xi_{2}<\xi_{3}$). The study of the coordinate
singularities at $x=\xi_{1}$ and $x=\xi_{2}$ leads to the
determination of $\xi_{F}$ as:
\begin{equation}
\xi_{F}=\frac{\xi_{1}\xi_{2}-\xi_{3}^{2}}{\xi_{1}+\xi_{2}-2\xi_{3}}
\in \left(\xi_{2},\xi_{3}\right).
\end{equation}
Emparan and Reall have established the asymptotically flat
character of (\ref{ERmetric}), as well as existence of an
analytic extension across an analytic  Killing horizon%
\footnote{We follow the terminology
of~\cite{ChCo}.\label{Ffterm}}
at $z=\xi_3$. The extension given in~\cite{EmparanReall} is
somewhat similar of the extension of the Schwarzschild metric
that one obtains by going to Eddington-Finkelstein coordinates,
and is \emph{not} maximal. Now, existence of an analytic
extension with a bifurcate horizon, \emph{\`a la}
Kruskal-Szekeres, is guaranteed from this by an analytic
version of the analysis of R\'acz and Wald in~\cite{RaczWald2}.
But the global properties of an extension so constructed are
not clear. It is therefore of interest to present an explicit
extension with good properties. This extension is constructed
in Section~\ref{Sbifhor}, and its global properties are studied
in the remaining sections.

As in~\cite{EmparanReall} we assume throughout that
\bel{xrange}
 \xi_{1} \le x \le \xi_{2}
 \;.
\ee
As discussed in~\cite{EmparanReall}, the extremities correspond
to a north and south pole of $S^2$, with a function $\theta$
defined by $d\mutheta = dx/\sqrt{G(x)}$ providing a latitude on
$S^2$, except for the limit $x-z\to 0$, $x\to \xi_1$, which
corresponds to an asymptotically flat region,
see~\cite{EmparanReall}; a  detailed proof of asymptotic
flatness can be found in~\cite{ChBeijing}.
The surface
``$\{z=\infty\}$" can be identified with ``$\{z=-\infty\}$" by
introducing a coordinate $Y=-1/z$, with the metric extending
analytically across $\{Y=0\}$, see~\cite{EmparanReall} for
details.

We will denote by $(\mcM_{I\cup I\!I},g)$  the space-time
constructed by Emparan and Reall, as outlined above, where the
coordinate $z$ runs then over $(\xi_F,\infty] \cup
[-\infty,\xi_1]$. We will denote by $(\mcM_I,g)$ the subset of
$(\mcM_{I\cup I\!I},g)$ in which the coordinate $z$ runs over
$(\xi_3,\infty] \cup [-\infty,\xi_1]$; see
Figure~\ref{FigureER1}.

Strictly speaking, in the  definitions of  $(\mcM_{I\cup
I\!I},g)$ and  $(\mcM_I,g)$ we should have used different
symbols for the metric $g$; we hope that this will not lead to
confusions.

\section{The extension}
 \label{Sbifhor}
We start by working in the range $z\in (\xi_3,\infty)$; there
we define new coordinates $w,v$ by the formulae
\newcommand{\myb}{b}%
\newcommand{\myc}{\xi_2}%
\newcommand{\hpsi}{\hat \psi}%
\begin{equation}
\label{1dv}
 dv=dt+\frac{\myb dz}{ (z-\xi_3)(z-\myc)} \;,
\end{equation}
\begin{eqnarray}
\label{1dw} dw=dt- \frac{\myb dz}{  (z-\xi_3)(z-\myc)}  \;,
\end{eqnarray}
where $\myb$ is a  constant to be chosen shortly. (Our
coordinates $v$ and $w$ are closely related to, but not
identical, to the coordinates $v$ and $w$ used
in~\cite{EmparanReall} when extending the metric through the
Killing horizon $z=\xi_{3}$). Similarly to the construction of
the extension of the Kerr metric
in~\cite{CarterKerr,BoyerLindquist}, we define a new angular
coordinate $\hat{\psi}$ by:
\begin{equation}
\label{psi} d\hat{\psi}=d\psi -\myalph dt \;,
\end{equation}
where $\myalph$ is a constant to be chosen later.
Let
\bel{defmygamma}
 \mygamma:=\frac{1}{A}\sqrt{\frac{\nu}{\xi_{F}}}
 \;.
\ee
Using (\ref{1dv})--(\ref{psi}), we obtain
\begin{eqnarray}
\label{12.4}
dt=\frac{1}{2}(dv+dw)\; , \\
 \label{12.5}
dz=\underbrace{\frac{(z-\xi_3)(z-\myc)}{2\myb}}_{=:H(z)/2} (dv-dw)\; , \\
d\psi=d\hat{\psi}+\frac{\myalph }{2}(dv+dw)\; ,
\end{eqnarray}
which leads to
\begin{equation}
\label{1gvv}
 g_{vv}=g_{ww}=-\frac{F(x)}{4F(z)}\Big(1+\myalph \mygamma (\xi_{1}-z)  \Big)^{2}
 -\frac{ F(x)F(z) }{4A^{2}(x-z)^{2} }
 \left(\frac{\myalph^{2} G(z) }{F(z)}
 + \frac{H^2(z)}{G(z)}
 \right)
  \; ,\end{equation}
\begin{equation}
\label{1gvw}
 g_{vw}=-\frac{F(x)}{4F(z)}\Big(1+\myalph \mygamma (\xi_{1}-z)  \Big)^{2}
 -\frac{ F(x)F(z) }{4A^{2}(x-z)^{2} }
 \left(\frac{\myalph^{2} G(z) }{F(z)}
 - \frac{H^2(z)}{G(z)}
 \right)
  \; ,\end{equation}
\begin{equation}
\label{1gvpsi}
g_{v\hat{\psi}}=g_{w\hat{\psi}}=-\frac{F(x)}{2F(z)}\mygamma(\xi_{1}-z)\Big(
1+\myalph \mygamma (\xi_{1}-z)\Big)
-\frac{F(x)G(z)\myalph}{2A^{2}(x-z)^{2}} \; ,\end{equation}
\begin{equation}
\label{1gpsipsi}
 g_{\hat{\psi}\hat{\psi}}=-\frac{F(x)}{F(z)}\mygamma^{2}(\xi_{1}-z)^{2}-\frac{F(x)G(z)}{A^{2}(x-z)^{2}}
 \; .
\end{equation}
The Jacobian of the coordinate transformation is
$$\frac{\partial(w,v,\hat{\psi},x,\varphi)}{\partial(t,z,\psi,x,\varphi)}
 =2 \frac{\partial v}{\partial z} =  \frac{2\myb}{(z-\xi_2)(z-\xi_3)}
  \;.
$$
In the original coordinates $(t,z,\psi,x,\varphi)$ the
determinant of $g$ was
\begin{equation}
\label{1detg}
 \det(g_{(t,z,\psi,x,\varphi)})=-\frac{F^{2}(x)F^{4}(z)}{A^{8}(x-z)^{8}}\;,
\end{equation}
so that in the new coordinates it reads
\begin{equation}
\label{1detgvw} \det(g_{(w,v,\hat{\psi},x,\varphi)})
 = -
 \frac{ F^{2}(x)F^{4}(z)(z-\xi_2)^2(z-\xi_3)^2}{4A^{8}\myb^2(x-z)^{8} }
\; .
\end{equation}
This last expression is negative on
$\left(\xi_{F},\infty\right)\setminus\{\xi_{3}\}$, and has a
second order zero at $z=\xi_{3}$. In order to remove this
degeneracy  one introduces
\begin{eqnarray}
 {\hat{v}=\exp(cv)}\;, \quad
 \hat{w}=-\exp(-cw)
  \;,
   \label{coordexp}
\end{eqnarray}
where $c$ is some constant to be chosen. Hence we have
\begin{eqnarray}
 \label{dhatv}
d\hat{v}=c\hat{v}dv\;, \quad d\hat{w}=-c\hat{w}dw
 \;,
\end{eqnarray}
and the determinant in the coordinates
$(\hat{w},\hat{v},\hat{\psi},x,\varphi)$ reads
\begin{equation}
\label{1detexp} \det(g_{(\hat{w},\hat{v},\hat{\psi},x,\varphi)})
 =-
 \frac{ F^{2}(x)F^{4}(z)(z-\xi_2)^2(z-\xi_3)^2}{4A^{8}\myb^2(x-z)^{8} c^{4}\hat{v}^{2}\hat{w}^{2}}  \;.
\end{equation}
But one has $\hat{v}^{2}\hat{w}^{2}=\exp(2c(v-w))$, so that
\begin{eqnarray}
\label{1machin}
\hat{v}^{2}\hat{w}^{2}
  &= &\exp\left(
{4c\myb} \int \frac{1}{(z-\xi_{2}) (z-\xi_{3})}dz\right)
\\
 &=&
      \exp\left(  \frac{4c\myb  }{ (\xi_3-\xi_{2}) }(
      \ln(z-\xi_{3})-
  \ln(z-\xi_{2}))\right)
 \; .\nonumber
\end{eqnarray}
Taking into account \eq{1machin}, and the determinant
(\ref{1detexp}), we choose  the constant $c$ to satisfy:
\bel{1choixc} \frac{2c\myb  }{ (\xi_3-\xi_{2}) }=1 \;.
 \ee
We obtain
\bel{1mydet}
 \hat v \hat w = - \frac
 { z-\xi_3 }
 { z-\xi_{2} }
\;, \ee
and
\begin{equation}
\label{1detexp2} \det(g_{(\hat{w},\hat{v},\hat{\psi},x,\varphi)})
 =-
 \frac{ F^{2}(x)F^{4}(z)(z-\xi_{2})^4}{4A^{8}\myb^2(x-z)^{8} c^{4} }  \;.
\end{equation}
With this choice, the determinant of $g$ in the
$(\hat{w},\hat{v},\hat{\psi},x,\varphi)$ coordinates extends to
a strictly negative analytic function on
$\{z\in\left(\xi_{F},\infty\right)\}$. In fact, $z$ is an
analytic function of $\hat{v}\hat{w}$ on $\{\hat{v} \hat{w}\ne
-1\}$ (that last set corresponds to $z=\infty \Leftrightarrow
Y=0$, we will return to this shortly):
\bel{1mydetx}
  z  =   \frac
 {  \xi_3 +\xi_{2}\hat v \hat w}
 {1+\hat v \hat w}
\;. \ee
 In the
$(\hat{w},\hat{v},\hat{\psi},x,\varphi)$ coordinates, one
obtains the coefficients of the metric from \eq{dhatv} using
\begin{equation}
\label{gfinal}
 g_{\hat{v}\hat{v}}=\frac{1}{c^{2}\hat{v}^{2}}g_{vv}\;,
  \quad
 g_{\hat{w}\hat{w}}=\frac{1}{c^{2}\hat{w}^{2}}g_{ww}\;,
 \end{equation}
\begin{equation}
g_{\hat{v}\hat{w}}=-\frac{1}{c^{2}\hat{v}\hat{w}}g_{vw}\;,
 \quad
g_{\hat{v}\hat{\psi}}=\frac{1}{c\hat{v}}g_{v\hat{\psi}}\;,\nonumber
 \quad
 g_{\hat{w}\hat{\psi}}=-\frac{1}{c\hat{w}}g_{w\hat{\psi}}\;.
\end{equation}

In order to show that the coefficients of the metric are
analytic on  the set
\bel{1domhatvhatw}
 \Big\{\hat{w},\hat{v}\ | \ z(\hat{v}\hat{w})
> \xi_{F}\Big\} =
 \Big\{\hat{w},\hat{v}\ | \ -1< \hat{v}\hat{w} < \frac
 { \xi_3-\xi_F }
 { \xi_F-\xi_{2} }
 \Big\}
\ee
it is convenient to  write
\begin{equation}
\label{gfinal2}
g_{\hat{v}\hat{v}}=\frac{1}{c^{2}\hat{v}^{2}\hat{w}^{2}}\hat{w}^{2}g_{vv}\;,
 \quad
 g_{\hat{w}\hat{w}}=\frac{1}{c^{2}\hat{v}^{2}\hat{w}^{2}}\hat{v}^{2}g_{ww}\;,
\end{equation}
\be
g_{\hat{v}\hat{w}}=-\frac{1}{c^{2}\hat{v}\hat{w}}g_{vw}\;,
 \quad
g_{\hat{v}\hat{\psi}}=\frac{1}{c\hat{v}\hat{w}}\hat{w}g_{v\hat{\psi}}\;,\nonumber
 \quad
g_{\hat{w}\hat{\psi}}=-\frac{1}{c\hat{v}\hat{w}}\hat{v}g_{w\hat{\psi}}\;.
\ee
Hence, to make sure that all the coefficients of metric are
well behaved at  $\{\hat{w},\hat{v} \in \mathbb{R}\ |\
z=\xi_{3}\}$ (i.e. $\hat{v}=0$ or $\hat{w}=0$), it suffices to
check that there is a multiplicative factor $(z-\xi_{3})^{2}$
in $g_{vv}=g_{ww}$, as well as a multiplicative factor
$(z-\xi_{3})$ in $g_{vw}$ and in
$g_{v\hat{\psi}}=g_{w\hat{\psi}}$. In view of
(\ref{1gvv})--(\ref{1gpsipsi}), one can see that this will be
the case if , first, $\myalph$ is chosen so that $1+\myalph
\mygamma(\xi_{1}-z)=\myalph \mygamma (\xi_{3}-z)$, that is
\bel{alpha} \myalph=\frac{1}{\mygamma(\xi_{3}-\xi_{1})}, \ee
and then, if $\myb$ is chosen such that
\bea
 0
 &=&
 -\frac{\myalph^{2} \nu \xi_F (\xi_3-\xi_1)    }{\xi_3-\xi_F}
 + \frac{  1  }{\nu \myb^2 (\xi_3-\xi_1)
 }
 \;.
\eeal{24XI8.2}
\Eq{24XI8.2} will hold if we set
\bea
 \myb^2
 &=&
  \frac{ (\xi_3-\xi_F)}{\nu^2  \myalph^{2}  \xi_F  (\xi_3-\xi_1)^2}
 \;.
\eeal{24XI8.3}

So far we have been focussing on the region
$z\in(\xi_F,\infty)$, which overlaps only with part of the
manifold ``$\{z\in (\xi_3,\infty]\cup[-\infty,\xi_1]\}$". A
well behaved coordinate on that last region is $Y=-1/z$. This
allows one to go smoothly through $Y=0$ in \eq{1mydet}:
\bel{1mydet2}
 \hat v \hat w = - \frac
 {  1+\xi_3 Y  }
 { 1+\xi_2 Y }
  \quad
   \Longleftrightarrow
   \quad
    Y  =  - \frac
 {1+\hat v \hat w}
 {  \xi_3 +\xi_2\hat v \hat w}
 \;.
\ee
In other words, $\hat v \hat w$ extends analytically to the
region of interest, $0\le Y \le -1/\xi_1$ (and in fact beyond,
but this is irrelevant to us). Similarly, the determinant $
\det(g_{(\hat{w},\hat{v},\hat{\psi},x,\varphi)})$ extends
analytically across $Y=0$, being the ratio of two polynomials
of order eight in $z$ (equivalently, in $Y$), with limit
\begin{equation}
\label{1detexp3} \det(g_{(\hat{w},\hat{v},\hat{\psi},x,\varphi)})
 \to_{z\to\infty}-
 \frac{ F^{2}(x) }{4A^{8}\myb^2 c^{4} \xi_F ^4}  \;.
\end{equation}
We conclude that the construction so far produces an analytic
Lorentzian metric on the set
\bel{1domhatvhatwx}
 \hat \Omega:= \Big\{\hat{w},\hat{v}\ | \  - \frac
 {   \xi_3 -\xi_1 }
 {  \xi_2 -\xi_1}\le  \hat{v}\hat{w} < \frac
 { \xi_3-\xi_F }
 { \xi_F-\xi_{2} }
 \Big\}\times S^1_{\hat \psi} \times S^2_{(x,\varphi)}
  \;,
 \ee
Here a subscript on $S^k$ points to the names of the
corresponding local variables.
\begin{figure}[ht]
\begin{center}
\hspace{-3cm}{
 \psfrag{zxidef}{\huge$ z=\xi_F$}
 \psfrag{zxidetrois}{\huge$ z=\xi_3$}
 \psfrag{mun}{\huge$ \mcM_I$}
 \psfrag{mdeux}{\huge$ \mcM_{I\!I}$}
 \psfrag{mtrois}{\huge$ \mcM_{I\!I\!I}$}
 \psfrag{mquatre}{\huge$ \mcM_{IV}$}
 \psfrag{scriun+}{\huge$ \scri_I^+$}
 \psfrag{scriun-}{\huge$ \scri_I^-$}
 \psfrag{scritrois+}{\huge$ \scri_{I\!I\!I}^+$}
 \psfrag{scritrois-}{\huge$ \scri_{I\!I\!I}^-$}
\hspace{2.5cm}\resizebox{3in}{!}{\includegraphics{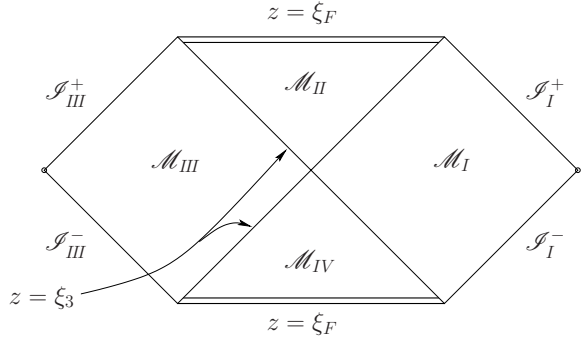}} }
\caption{$\hmcM$ with its various subsets. For example, $\mcM_{I\cup
I\!I}$ is the union of $ \mcM_{I }$ and of $ \mcM_{ II}$ and of that
part of $\{z=\xi_3\}$ which lies in the intersection of their
closures; this is the manifold constructed in~\cite{EmparanReall}.
Very roughly speaking, the various $\scri$'s correspond to
$x=z=\xi_1$. It should be stressed that this is {neither} a
\emph{conformal} diagram, nor is the space-time a product of the
figure times $S^2\times S^1$: {{$\mcM_I$ cannot be the product of
the depicted diamond with $S^2\times S^1$, as this product is not
simply connected, while $\mcM_I$ is}}. But the diagram represents
accurately the causal relations between the various $\mcM_{N}$'s, as
well as the geometry near the bifurcate horizon {$z=\xi_F$},  as the
manifold \emph{does have} a product structure there.}
\label{FigureER1}
\end{center}
\end{figure}

The map
\bel{ERiso}
 (\hat{w},\hat{v}, \hat \psi,x,\varphi)\mapsto(-\hat w,-\hat
 v,-\hat \psi,x, -\varphi)
\ee
is an orientation-preserving analytic isometry of the
analytically extended metric on $\hat \Omega$. It follows that
the manifold
$$
 \hmcM
$$
obtained by gluing together $\hat \Omega$ and  two isometric
copies of $(\mcM_I,g)$ can be equipped with the obvious
Lorentzian metric, still be denoted by $g$, which is
furthermore analytic. The second copy of $(\mcM_I,g)$ will be
denoted by $(\mcM_{I\!I\!I},g)$; compare
Figure~\ref{FigureER1}. The reader should keep in mind the
polar character of the coordinates around the relevant axes of
rotation, and the special character of the ``point at infinity"
$z=\xi_1=x$.

\section{Global structure}
\subsection{The event horizon has $S^2\times S^1\times \R$
topology}

As shortly reviewed in Section~\ref{SERm}, it is shown
in~\cite{EmparanReall} how to extend  the metric \eq{ERmetric}
across
$$
 \mcE:=\{z=\xi_3\}
$$
to an analytic metric on $\mcM_{I\cup I\!I}$. Further, we have
shown in Section~\ref{Sbifhor} how to extend $g$ to an analytic
metric on $\hmcM$. Now,
\bel{gupzz} g(\nabla z, \nabla z)= g^{zz} = -\frac{A^2 (x-z)^2
G(z)}{F(x)F(z)}
 \ee
in the region $\{z>\oldxifour \}$, and by analyticity  this
equation remains valid on $\{z>\xi_F \}$. Equation~\eq{gupzz}
shows that $\mcE$ is a null hypersurface, with $z$ being a time
function on $\{\xi_F<z< \oldxifour \}$. The usual choice of
time orientation implies that $z$ is strictly decreasing along
future directed causal curves in the region $\{\hat v>0\;, \hat
w>0\}$, and strictly increasing along such curves in the region
$\{\hat v<0\;, \hat w<0\}$.
In particular no causal future directed curve can leave the
region $\{\hat v>0\;, \hat w>0\}$. Hence the space-time
contains a black hole region.

However, it is not clear that $\mcE$ is the  \emph{event
horizon} within the Emparan-Reall space-time $( \mcM_{I\cup
I\!I},g)$, because the actual event horizon could be enclosing
the region $z<\xi_3$. To show that this is not the case,
consider the ``area function", defined as the determinant, say
$W$, of the matrix
$$
 g(K_i,K_j)\;,
$$
where the $K_i$'s, $i=1,2,3$, are the Killing vectors equal to
$\partial_t$, $\partial_\psi$, and $\partial_\varphi$ in the
asymptotically flat region. In the original coordinates of
\eq{metric20} this equals
\bel{Wdet} \frac{F(x)G(x)F(z)G(z)}{ A^{4} (x-z)^{4}}
 \;.
\ee
Analyticity implies that this formula is valid throughout
$\mcM_{I\cup I\!I}$, as well as $\hmcM$. Now,
$$F(z)G(z)= \frac \nu {\xi_F}
(\xi_F-z)(z-\xi_1)(z-\xi_2)(z-\xi_3)
 \;,
$$
and, in view of the range \eq{xrange} of the variable $x$, the
sign of \eq{Wdet} depends only upon the values of $z$. Since
$F(z)G(z)$ behaves as $- \nu z^4/\xi_F$ for large $z$, $W$ is
negative both for $z<\xi_1$ and for $z>\xi_3$. Hence, at each
point $p$ of those two regions the set of vectors in $T_p\mcM$
spanned by the Killing vectors is timelike. So, suppose for
contradiction, that the event horizon $\mcH$ intersects the
region $\{ z\in(\xi_3,\infty]\}\cup\{z\in [-\infty,\xi_1)\}$;
here ``$z=\pm \infty$" should be understood as $Y=0$, as
already mentioned in the introduction. Since $\mcH$ is a null
hypersurface invariant under isometries, every Killing vector
is tangent to $\mcH$. However, at each point at which $W$ is
negative there  exists a linear combination of the Killing
vectors which is timelike. This gives a contradiction because
no timelike vectors are tangent to a null hypersurface.

We conclude that $\{z=\xi_3\}$ forms indeed the event horizon
in the space-time $(\mcM_{I\cup I\!I},g)$ (as defined at the
end of Section~\ref{SERm}), with topology $\R\times S^1 \times
S^2$.

The argument just given also shows that the domain of outer
communications within $(\mcM_I,g)$ coincides with $(\mcM_I,g)$.

Similarly, one finds that the domain of outer communications
within $(\hmcM,g)$, or that within $(\mcM_{I\cup I\!I},g)$,
associated with an asymptotic region lying in $(\mcM_I,g)$, is
$(\mcM_I,g)$. The boundary of the d.o.c.\ in $(\hmcM,g)$ is a
subset of the set $\{z=\xi_3\}$, which can be found by
inspection of Figure~\ref{FigureER1}.

\subsection{Inextendibility at $z=\xi_F$}
 \label{ssIxif}

The obvious place where $(\hmcM,g)$ could be enlarged is at
$z=\xi_F$. To show that no extension is possible there,
consider%
\footnote{This inextendibility criterion has been introduced
in~\cite{BCIM} (see the second part of Proposition~5, p.~139
there).}
the norm of the Killing vector
field $\partial_t$:
\bel{gtbl}
 g(\partial_t,\partial_t) = -\frac{F(x)}{F(z)} \to_{\xi_F<z \to
 \xi_F}\infty \quad \mbox{(recall that $F(x)\ge 1-\frac{\xi_2}{\xi_F}>0$).}
\ee
 Suppose,
for contradiction, that there exists a $C^2$ extension of the
metric through $\{z=\xi_F\}$. Recall that any Killing vector
field $X$ satisfies the set of equations
\bel{Kileqx}
 \nabla_\alpha \nabla_\beta X_\sigma = R_{\lambda \alpha \beta \sigma} X^\lambda
 \;.
\ee
But the overdetermined set of linear equations \eq{Kileqx}
together with existence of a $C^2$ extension implies that
$\partial_t$ extends, in $C^2$, to $\{z=\xi_F\}$, contradicting
\eq{gtbl}.

An alternative way, demanding somewhat more work, of proving
that the Emparan-Reall metric is $C^2$--inextendible across
$\{z=\xi_F\}$, is to notice that
$R_{\alpha\beta\gamma\delta}R^{\alpha\beta\gamma\delta}$ is
unbounded along any curve along which $z$ approaches $\xi_F$.
This has been pointed out to us by Harvey Reall (private
communication), and has been further verified by Alfonso
Garcia-Parrado and Jos\'e Mar\'ia Mart\'in Garc\'ia using the
symbolic algebra package {\sc xAct}~\cite{xAct}:
\bel{KrER}
 R_{\alpha\beta\gamma\delta}R^{\alpha\beta\gamma\delta}
 = \frac{12 A^4 \xi_F^4 G(\xi_F)^2 (x-z)^4\left(1+O(z-\xi_F)\right)}{(\xi_F-x)^2
(z-\xi_F)^6}
 \;.
\ee
We are grateful to Alfonso and Jos\'e Mar\'ia for carrying out
the calculation.

\subsection{Maximality}
 \label{sMax}

Let $k\in \R\cup\{\infty\}\cup \{\omega\}$. The
$(n+1)$--dimensional space-time $(\tmcM,\tg)$ is said to be
\emph{a $C^k$--extension} of an $(n+1)$--dimensional space-time
$(\mcM,g)$ if there exists a $C^k$--immersion $\psi: \mcM\to
\tmcM$ such that $\psi^* \tg = g$, and such that $\psi(\mcM)\ne
\tmcM$. A space-time $(\mcM,g)$ is said to be
\emph{$C^k$-maximal}, or \emph{$C^k$-inextendible}, if no
$C^k$--extensions of $(\mcM,g)$ exist.

A \emph{scalar invariant} is a function which can be calculated
using the geometric objects at hand and which is invariant
under isometries. For instance, a function $\alpha_g$ which can
be calculated in local coordinates from the metric $g$ and its
derivatives will be a scalar invariant if, for any local
diffeomorphism $\psi$ we have
\bel{scinv}
 \alpha_g (p) = \alpha_{\psi^*g}(\psi^{-1}(p))
 \;.
\ee
In the application of our Theorem~\ref{Tscompl} below to the
Emparan-Reall space-time one can use the scalar invariant
$g(X,X)$, calculated using a metric $g$ and a Killing vector
$X$. In this case the invariance property \eq{scinv} reads
instead
\bel{scinv2}
 \alpha_{g,X} (p) = \alpha_{\psi^*g, (\psi^{-1})^*X}(\psi^{-1}(p))
 \;.
\ee

A scalar invariant $f$ on $(\mcM,g)$ will be called a
\emph{$C^k$\obstruction{} scalar} if $f$ satisfies the
following property: For every $C^k$--extension $(\tmcM,\tg)$ of
$(\mcM,g)$ and for any bounded timelike geodesic segment
$\gamma$ in $\mcM$ such that $\psi(\gamma)$ accumulates at the
boundary $\partial(\psi(\mcM))$ (where $\psi$ is the immersion
map $\psi:\mcM\to \tmcM$), the function $f$ is bounded along
$\gamma$.

An example of a $C^2$\obstruction{} scalar is the Kretschmann
scalar, which writes
$R_{\alpha\beta\gamma\delta}R^{\alpha\beta\gamma\delta}$. As
explained in Section~\ref{ssIxif}, another example is provided
by the norm $g(X,X)$ of a Killing vector $X$ of $g$. Any
constant function is a compatibility scalar in this
terminology, albeit not very useful in practice.

We shall need the following generalisation of a maximality
criterion of~\cite[Appendix~C]{SCC}:%
\footnote{JC acknowledges useful discussions with M.~Herzlich
concerning the problem at hand.}

\begin{Proposition}
 \label{Pmax}
 Let $k\ge 2$.
  Suppose that  every timelike
 geodesic $\gamma$ in
$(\mcM,g)$ is either complete, or some $C^k$\obstruction{}
scalar is unbounded on $\gamma$. Then $(\mcM,g)$ is
$C^k$-inextendible.
\end{Proposition}

\proof
Suppose that there exists a $C^k$--extension $(\tmcM,\tg)$ of
$(\mcM,g)$, with immersion $\psi:\mcM\to \tmcM$. We identify
$\mcM$ with its image $\psi(\mcM)$ in $\tmcM$.

Let $p\in \partial \mcM$ and let $\mcO$ be a globally
hyperbolic neighborhood of $p$. Let $q_n\in \mcM$ be a sequence
of points approaching $p$, thus $q_n \in \mcO$ for $n$ large
enough. Suppose, first, that there exists $n$ such that $q_n
\in I^+(p)\cup I^-(p)$. By global hyperbolicity of $\mcO$ there
exists a timelike geodesic segment $\gamma$ from  $q_n$ to $p$.
Then the part of $\gamma$ which lies within $\mcM$ is
inextendible within $\mcM$ and has finite affine length.
Furthermore every $C^k$\obstruction{} scalar is bounded on
$\gamma$. But there are no such geodesics through $q_n$ by
hypothesis. We conclude that
\bel{empint}
 (I^+(p)\cup I^-(p))\cap \mcM = \emptyset
 \;.
\ee

Let $q\in (I^+(p)\cup I^-(p))\cap \mcO$, thus $q\not\in \mcM$
by \eq{empint}. Since $I^+(q)\cup I^-(q)$ is open, and $p\in
I^+(q)\cup I^-(q)$, we have $q_n \in I^+(q)\cup I^-(q)$ for all
$n$ sufficiently large, say $n\ge n_0$. Let $\gamma$ be a
timelike geodesic segment from $q_{n_0}$ to $q$. Since $q$ is
not in $\mcM$, the part of $\gamma$ that lies within $\mcM$ is
inextendible within $\mcM$ and has finite affine length, with
all $C^k$\obstruction{} scalars bounded. This is again
incompatible with our hypotheses, and the result is
established.
\qed

\medskip

In Section~\ref{sGeod} below (see Theorem~\ref{Tglobcontge}) we
show that all maximally extended causal geodesics of our
extension $(\hmcM,\mhatg)$ of the Emparan-Reall space-time are
either complete, or reach the singular boundary $\{z=\xi_F\}$
in finite affine time. This, together with Section~\ref{ssIxif}
and Proposition~\ref{Pmax} gives:

\begin{Theorem}
 \label{TMLIm}
$(\hmcM,\mhatg)$ is maximal within the class of $C^2$
Lorentzian manifolds.
\end{Theorem}

\subsection{Global hyperbolicity}
 \label{sGh}

In this section we  show that $(\hmcM,g)$ is globally
hyperbolic. We shall need the following standard fact (see,
e.g.,~\cite[Lemma~13, p.408]{ONeill83}):

\begin{Lemma}
 \label{scausal}
Let $\alpha$ be a maximally extended causal curve in a strongly
causal space-time $(\mcM,g)$ meeting a compact set $K$. Then
$\alpha$ eventually leaves $K$,       never to return, both to
the future and to the past.
\end{Lemma}

Recall that in the region $\mcM _I$, the time-coordinate $t$ is
a time-function since $\nabla t$ is timelike; hence $\dot{t}>0$
along any future-directed causal curve. In terms of the
coordinates of Section~\ref{Sbifhor} we have:
\be
 t=\frac{1}{2c}\ln\left(-\frac{\hat{v}}{\hat{w}}\right)\;.
\ee
Letting $\hat v$ and $\hat w$ be the global coordinates of
Section~\ref{Sbifhor}, we define the hypersurface
\bel{hypdefJC}
 \hyp:=\{\hat{v}+\hat{w}=0\}\;.
\ee
(Thus, $\hyp$ extends smoothly  the hypersurface
$\{t=0\}\subset \mcM_I$, across the bifurcation surface $\{\hat
w=\hat v =0\}$, to its image in $\mcM_{I\!I\!I}$ under the map
\eq{ERiso}.) The hypersurface $\hyp $ is spacelike, and we wish
to show that it is Cauchy. We start by noting that it is
achronal: Indeed, on $\mcM_I$ the function $t$, as well as its
mirror counterpart on $\mcM_{I\!I\!I}$, are time functions on
$\mcM_{I }\cup\mcM_{I\!I\!I}$, so any connected causal curve
through those regions can meet $\hyp$ at most once. Next, since
$z$ is a time function on $\mcM_{I\!I }\cup\mcM_{IV}$,   any
causal curve entering $\mcM_{I\!I }\cup\mcM_{IV}$ from $\mcM_{I
}\cup\mcM_{I\!I\!I}$ cannot leave $\mcM_{I\!I }\cup\mcM_{IV}$,
and therefore cannot intersect again. A similar argument
applies to those causal curves which enter $\mcM_{I\!I }$ from
$\mcM_{IV}$, or vice-versa, and achronality follows.

So, from \cite[Property 6]{GerochDoD}, since $\hyp $ is closed,
spacelike, and achronal, it suffices to prove that every
inextendible null geodesic in $\hmcM$ intersects $\hyp $. For
this, let $\gamma$ be such a geodesic, say future-directed.

\begin{figure}[ht]
\begin{center}{
 \psfrag{hatweps}{   $\!\!\!\!\!\!\!\!\!\!\!\!\!\!\hat w=-\varepsilon_1$}
 \psfrag{hatw}{   $\!\hat w$}
 \psfrag{hatv}{   $\!\!\hat v$}
 \psfrag{coco}{   $ \hat v/\hat w=\hat v(s_1)/\hat w(s_1)$}
 \psfrag{hatveps}{   $\!\!\!\!\!\!\hat v= \varepsilon_1$}
 \psfrag{hatvhaw}{   $\!\!\hat v=-\hat w$}
 \psfrag{Kepsilon}{   $\!\!\!K_{\varepsilon_1}$}
 \resizebox{3in}{!}{\includegraphics{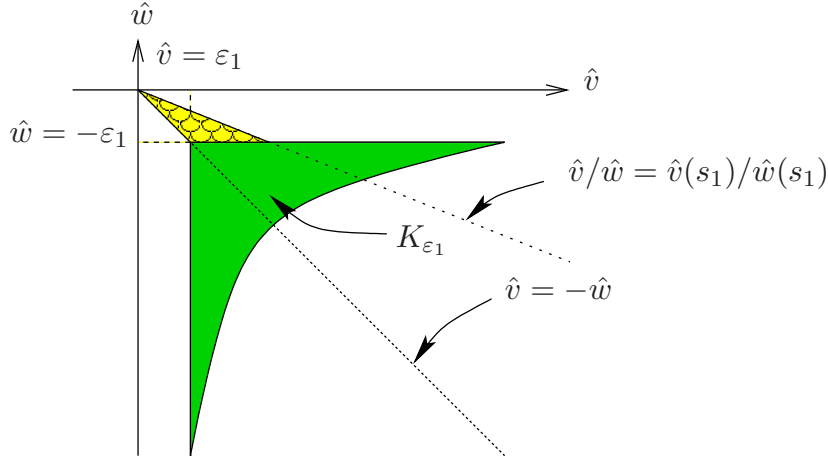}}
}
\caption{The set $K_{\varepsilon_1}$. Here $\hat w$ and $\hat v$ are the global coordinates of Section~\ref{Sbifhor}.
\label{Fglobhyp}}
\end{center}
\end{figure}

Suppose, first, that $\gamma(0)\in\mcM_I$; the argument in
$\mcM_{I\!I\!I}$ follows from isometry invariance. Choose
$\varepsilon
>0$ sufficiently small so that the set $K_\varepsilon$, defined
using the coordinates $\hat v,\hat w$ of
Section~\ref{Sbifhor}",
\bel{Kvarepsilon} K_\varepsilon:=\left\{\hat{v}\geq \varepsilon
,\ \hat{w}\leq -\varepsilon,\ \hat{v}\hat{w}\geq
 - \frac
 {   \xi_3 -\xi_1 }
 {  \xi_2 -\xi_1}+\varepsilon \right\} \times S^1 \times S^2
 \;,
\ee
contains $\gamma(0)$ (see Figure~\ref{Fglobhyp}).
Suppose, moreover, that $t(\gamma(0))>0$. Since $K_\varepsilon$
is  compact, and $(\mcM_I,g)$ is strongly causal (as it has a
time function), Lemma~\ref{scausal} implies that $\gamma$ must
leave $K_\varepsilon$, and never reenter, except perhaps after
exiting from $(\mcM_I,g)$ and returning again. But we have
already seen that $\gamma$ cannot return to $\mcM_I$ once it
exited, so $\gamma$ will indeed eventually never reenter
$K_\epsilon$. Choose any sequence $\varepsilon_i$ tending to
zero, then there exists a decreasing sequence $s_i$ such that
$\gamma$ leaves $K_{\varepsilon_i}$ at $\gamma(s_i)$, and never
renters.

Now, the ratio $ {-\hat{v}}/{\hat{w}}$ is a time function on
$\mcM_I$, and therefore increasing along $\gamma$ to the
future, and decreasing to the past. If $\gamma$  crosses the
hypersurface $\{ {-\hat{v}}/{\hat{w}}=1\}$ we are done. So
suppose that $\gamma$ does not. Then:

\begin{enumerate}
  \item Either $\gamma$ definitively leaves $\partial
      K_{\varepsilon_1}$ at $s=s_1$  through that part of
      the boundary $\partial K_{\varepsilon_1}$ on which
      $\hat w = -\varepsilon_1$. Then $\gamma$ remains in
      the tiled yellow triangle in Figure~\ref{Fglobhyp}
      until it leaves $\mcM_I$, and thus all subsequent
      definitive exit points $\gamma(s_i)\in \partial
      K_{\epsilon_i}$ lie on $\hat w = -\varepsilon_i$. So
      $\hat w(s_i)\to 0$ and $\hat v(s_i)\to 0$. The
      argument around equation \eq{15X.6} below shows that
      $\hat w(s )\to 0$ and $\hat v(s )\to 0$ in finite
      time.   By the analysis of Section~\ref{sGeod},
      $\gamma$ smoothly extends through
      $\{\hat{v}=\hat{w}=0\}$. But this is impossible since
      $\gamma$ is null,  while the only null geodesics
      meeting $\{\hat{v}=\hat{w}=0\}$ are the generators of
      the Killing horizon, entirely contained within
      $\{\hat{v}=\hat{w}=0\}$.
  \item Or $\gamma$ definitively leaves $\partial
      K_{\varepsilon_1}$ at $s=s_1$  through that part of
      the boundary $\partial K_{\varepsilon_1}$ on which
      $$\hat w \hat v
      = - \frac
 {   \xi_3 -\xi_1 }
 {  \xi_2 -\xi_1}+\varepsilon_1
 \;.
 $$
 From what has been said, all subsequent definitive exit
      points take place on
      $$\hat w \hat v
      = - \frac
 {   \xi_3 -\xi_1 }
 {  \xi_2 -\xi_1}+\varepsilon_i
 \;.
 $$
 This implies that $z(s) \rightarrow \xi_1$. We need to
      consider two cases:

       Suppose, first, that the constant of motion $c_\psi
      $ vanishes. Then $c_t=0$ is not possible
 by~\eq{lambda2} (recall that $\lambda=0$), while
 from~\eq{dott}
  \be \dot{t}(s ) \ \mbox{approaches} \
  -\frac{F(\xi_1)}{F(x(s))}c_t\;,
   \ee
  which is strictly bounded away from zero. So $\dot{t} $
  is positive and uniformly bounded away from zero
  sufficiently far to the past. This, together with
  Theorem~\ref{Tgeoddoc}, shows that $\gamma$  crosses $\{
  {-\hat{v}}={\hat{w}}\}\subset \hyp$,  which contradicts
  our assumption.

   It remains to consider the possibility that  $c_\psi
  \neq 0$. \Eq{lambda2} implies that $x(s) \rightarrow
  \xi_1$. Hence $\gamma$ enters, and remains, within the
asymptotically flat region. There $\gamma$ is forced to
  cross $\{\frac{-\hat{v}}{\hat{w}}=1\}$ by arguments known
  in principle, contradicting again our assumption.
\end{enumerate}

We conclude that $\gamma$'s intersecting $\mcM_I$ or
$\mcM_{I\!I\!I}$, with $t(\gamma(0))>0$, meet $\hyp$ when
followed to the past.

The proof that future directed causal geodesics $\gamma$
intersecting $\mcM_I$ or $\mcM_{I\!I\!I}$, with
$t(\gamma(0))<0$, meet $\hyp$, is identical: one needs instead
to follow $\gamma$ to the future rather than to the past.
Alternatively one can invoke the existence of   isometries of
those regions which map $t$ to $-t$ and $\psi$ to $-\psi$.

Suppose, finally, that $\gamma(0)\in \mcM_{I\!I}\cup
\mcM_{I\!V}$. By Proposition~\ref{Preachbif}, the null geodesic
$\gamma$ exits in finite time through the bifurcate horizon
$\{\hat w \hat v =0\}$. If $\gamma$ exits through  $\{\hat w
=\hat v =0\}$ it intersects $\hyp$ there, and we are done;
otherwise it enters $\mcM_{I}\cup \mcM_{I\!I\!I}$. But we have
just seen that $\gamma$ must then intersect $\hyp$, and the
proof of global hyperbolicity  is complete.

\subsection{$\scri$}
 \label{sScri}

In this section we address the question of existence of
conformal completions at null infinity \emph{\`a la Penrose},
for a class of higher dimensional stationary space-times that
includes the Emparan-Reall metrics; see the Appendix
in~\cite{Damour:schmidt} for the $3+1$ dimensional case.

We start by noting that any stationary asymptotically flat
space-time which is vacuum, or electro-vacuum, outside of a
spatially compact set is necessarily \emph{asymptotically
Schwarzschildian}, in the sense that there exists a coordinate
system in which the leading order terms of the metric have the
Schwarzschild form, with the error terms falling-off one power
of $r$ faster:
\bel{SchwAs}
 g= g_m + O(r^{-(n-1)})\;,\quad
\ee
where $g_m$ is the Schwarzschild metric of mass $m$, and the
size of the decay of the error terms in \eq{SchwAs} is measured
in a manifestly asymptotically Minkowskian coordinate system.
The proof of this fact is outlined briefly
in~\cite[Section~2]{ChBeig3}. In that last reference it is also
shown that the remainder term has a full asymptotic expansion
in terms of inverse powers of $r$ in dimension $2k+1$, $k\ge
3$, or in dimension $4+1$ for static metrics. Otherwise, the
remainder is known to have an asymptotic expansion in terms of
inverse powers of $r$ and of $\ln r$, and whether or not there
will be non-trivial logarithmic terms in the expansion is not
known in general.

In higher dimensions, the question of existence of a conformal
completion at null infinity is straightforward:  We start by
writing the $(n+1)$--dimensional Minkowski metric as
\bea \label{genER} &\eta=-  dt^{2}+ dr^{2}+r^2 h\;,& \eea
where $h$ is the round unit metric on an $(n-2)$-dimensional
sphere. Replacing $t$ by the standard retarded time $ u=t-r $,
one is led to the following form of the metric $g$:
\bel{genhatco2ER} g =-   d u^2 -2du \,dr +r^2 h + O(r^{-(n-2)})
dx^\mu dx^\nu
 \;,
\ee
where the $dx^\mu$'s are the manifestly Minkowskian coordinates
$(t,x^1,\ldots , x^n) $ coordinates for $\eta$. Setting $x=1/r$
in \eq{genhatco2ER} one obtains
\bel{genhatco2ERx} g =\frac1 {x^2}\Big(-   x^2 d u^2 + 2du \,dx
+ h +O(x^{n-4})dy^\alpha dy^\beta\Big)
 \;,
\ee
with correction terms in \eq{genhatco2ERx} which will extend
smoothly to $x=0$ in the coordinate system $(y^\mu) =
(u,x,v^A)$, where the $v^A$'s are local coordinates on
$S^{n-2}$. For example, a term $O(r^{-2} ) dx^i dx^j$ in $g$
will  contribute a term
$$
 O(r^{-2})  dr^2 =O(r^{-2}) x^{-4} dx^2 = x^{-2} (O(1)dx^2)\;,
$$
which is bounded up to $x=0$ after a rescaling by $x^{ 2}$. The
remaining terms in  \eq{genhatco2ERx} are analyzed similarly.

In dimension $4+1$, care has to be taken to make sure that the
correction terms do not affect the signature of the metric so
extended; in higher dimension this is already apparent from
\eq{genhatco2ERx}.

So, to construct a conformal completion at null infinity for
the Emparan Reall metric it suffices to verify that the
determinant of the conformally rescaled metric, when expressed
in the coordinates described above, does not vanish at $x=0$.
This is indeed the case, and can be seen by calculating the
Jacobian of the map
$$
 (t,z,\psi,x,\varphi)\mapsto (u,x,v^A)\;;
$$
the result can then be used to calculate the determinant of the
metric in the new coordinates, making use of the formula for
the determinant of the metric in the original coordinates.

For a general stationary vacuum $4+1$ dimensional metric one
can always transform to the coordinates, alluded to above, in
which the metric is manifestly Schwarzschildian in leading
order. Instead of using $(u=t-r,x=1/r)$ one can use coordinates
$(u_m,x=1/r)$, where $u_m$ is the corresponding null coordinate
$u$ for the $4+1$ dimensional Schwarzschild metric. This will
lead to a conformally rescaled metric with the correct
signature on the conformal boundary. Note, however, that this
transformation might introduce log terms in the metric, even if
there were none to start with; this is why we did not use this
above.

In summary, whenever a stationary, vacuum, asymptotically flat,
$(n+1)$--dimensional metric, $4 \ne n\ge 3$, has an asymptotic
expansion in terms of inverse powers of $r$, one is led to a
smooth $\scri$. This is the case for any such metric in
dimensions $3+1$ or $2k+1$, $k\ge 3$. In the remaining
dimensions one always has a polyhomogeneous conformal
completion at null infinity, with a conformally rescaled metric
which is $C^{n-4}$ up-to-boundary. For the Emparan-Reall metric
there exists a completion which has no logarithmic terms, and
is thus $C^\infty$ up-to-boundary.

\medskip

\subsection{Uniqueness?}
 \label{sU}

\subsubsection{Distinct extensions}
 \label{sssDe}

We start by noting that maximal analytic extensions of
manifolds are not unique. The simplest counterexamples are as
follows: remove a subset $\Omega$ from a maximally extended
manifold $\mcM$ so that $\mcM\setminus \Omega$ is not simply
connected, and pass to the universal cover; extend maximally
the space-time so obtained, if further needed. This provides a
new maximal extension. Whether or not such constructions can be
used to classify all maximal analytic extensions remains to be
seen.

One can likewise ask the question, whether it is true that
$(\hmcM,\mhatg)$ is unique within the class of simply connected
analytic extensions of $(\mcM_I,g)$ which are inextendible and
globally hyperbolic. The following variation of the
construction gives a negative answer, when ``inextendible" is
meant as ``inextendible within the class of globally hyperbolic
manifolds": Let $\hyp$ be  the Cauchy surface $\{t=0\}$ in
$\hmcM$, as described in Section~\ref{sGh} (see \eq{hypdefJC}),
and remove from $\hyp$ a closed subset $\Omega$ so that
$\hyp\setminus \Omega$ is not simply connected. Let $\thyp$ be
a maximal analytic extension of the universal covering space of
$\hyp\setminus \Omega$, with the obvious Cauchy data inherited
from $\hyp$, and let $(\tmcM,\tg)$ be the maximal globally
hyperbolic development thereof. Then $(\tmcM,\tg)$ is a
globally hyperbolic analytic extension of $(\mcM_I,g)$ which is
maximal in the class of globally hyperbolic manifolds, and
distinct from $(\hmcM,g)$.

The examples just discussed  will exhibit the following
undesirable feature: existence of maximally extended geodesics
of finite affine length near which the space-time is locally
extendible in the sense of \cite{Racz}. This does not happen in
$(\hmcM,g)$. It turns out that there exists at least one more
maximal extension of the Emparan-Reall space-time $(\mcM_I,g)$
which does not suffer from this \emph{local extendibility}
pathology. This results from a general construction which
proceeds as follows:

\newcommand{\psimap}{\Psi}%
Consider any spacetime $(\mcM,g)$, and let $\mcM_I$ be an open
subset of $\mcM$. Suppose that there exists an isometry
$\psimap $ of $(\mcM,g)$ satisfying: a) $\psimap$ has no fixed
points; b) $\psimap (\mcM_I)\cap \mcM_I=\emptyset$; and c)
$\psimap ^2$ is the identity map. Then, by a) and c),
$\mcM/\psimap $ equipped with the obvious metric (still denoted
by $g$) is a Lorentzian manifold. Furthermore, by b), $\mcM_I$
embeds diffeomorphically into $\mcM/\psimap$ in the obvious
way. It follows from the results in~\cite{Nomizu} that
$(\mcM/\psimap, g)$ is analytic if  $(\mcM,g)$ was
(compare~\cite[Appendix~A]{ChAscona}).

Keeping in mind that a space-time is time-oriented by
definition, $\mcM/\psimap $ will be a space-time if and only if
$\psimap $  preserves time-orientation. If $\mcM$ is simply
connected, then $\pi_1(\mcM/\psimap)=\Z_2$.

As an example of this construction, consider the
Kruskal-Szekeres extension $(\mcM,g)$ of the Schwarzschild
space-time  $(\mcM_I,g)$; by the latter we mean a connected
component of the set $\{r>2m\}$ within $\mcM$. Let $(T,X)$ the
global coordinates on $\mcM$ as defined  on p.~153 of
\cite{Wald:book}. Let $\mathring \psimap :S^2\to S^2$ be the
antipodal map. Consider the four isometries $\psimap_{\pm\pm}$
of the Kruskal-Szekeres space-time defined by the formula, for
$p\in S^2$,
$$
 \psimap _{\pm\pm}\big(T,X,p\big)= \big(\pm T,\pm X,\mathring \psimap(p)\big)
 \;.
$$
Set $\mcM_{\pm\pm}:= \mcM/\psimap _{\pm\pm}$. As $\psimap_{++}$
is the identity in $(T,X)$, $\mcM_{++}$ is locally
asymptotically flat but is not asymptotically flat in the usual
sense, and thus irrelevant in our context.
 Next, both $\mcM_{-,\pm}$ are smooth maximal analytic
Lorentzian extensions of $(\mcM_I,g)$, but are not space-times.
Finally, $\mcM_{+-}$ is a maximal globally hyperbolic analytic
extension of the Schwarzschild manifold distinct from $\mcM$.
This is the ``$\R\mathbb{P}^3$ geon", discussed
in~\cite{FriedmanSchleichWitt}.

Similar examples can be constructed for the black ring
solution; we restrict attention to orientation, and
time-orientation, preserving maps. So, let $(\hmcM,\mhatg)$ be
our extension, as constructed above, of the domain of outer
communication $(\mcM_{I },g)$ within the Emparan-Reall
space-time $(\mcM_{I \cup I\!I},g)$, and let
$\psimap:\hmcM\to\hmcM $ be defined as
\bel{psimap}
 \psimap (\hat v,\hat w, \hat \psi, x, \varphi) = (\hat w,\hat v, \hat \psi+ \pi, x, -\varphi)
 \;.
\ee
By inspection of \eq{1gvv}-\eq{1gvpsi} and \eq{gfinal}, the map
$\psimap $ is an isometry, and clearly satisfies conditions a),
b) and c) above. Then $\hmcM/\psimap$ is a maximal, orientable,
time-orientable, analytic extension of $\mcM_{I }$ distinct
from $\hmcM$.

\subsubsection{A uniqueness theorem}

Our aim in this section is to prove a uniqueness result for our
extension $(\hmcM,\mhatg)$ of the Emparan-Reall space-time
$(\mcM_I,g)$. The examples of the previous section show that
the hypotheses are optimal:

\begin{Theorem}
 \label{TMLIu}
$(\hmcM,\mhatg)$ is unique within the class of simply connected
analytic extensions of $(\mcM_I,g)$ which have the property
that all maximally extended causal geodesics on which
$R_{\alpha\beta\gamma\delta}R^{\alpha\beta\gamma\delta}$ is
bounded are complete.
\end{Theorem}

Uniqueness is understood up to isometry. Theorem~\ref{TMLIu}
follows immediately from Theorem~\ref{Tscompl}, which we are
about to prove, and from our analysis of causal geodesics of
$(\hmcM,\mhatg)$ in Section~\ref{sGeod} below, see
Theorem~\ref{Tglobcontge} there.

For the record we state the corresponding result for the
Schwarzschild space-time, with identical (but simpler, as in
this case the geodesics are simpler to analyse) proof:

\begin{Theorem}
 \label{TMKSS} The Kruskal-Szekeres space-time is the unique
extension, within the class of simply connected analytic
extensions of the Schwarzschild region $r>2m$, with the
property that all maximally extended causal geodesics on which
$R_{\alpha\beta\gamma\delta}R^{\alpha\beta\gamma\delta}$ is
bounded are complete.
%
\end{Theorem}
We continue with some terminology. A maximally extended
geodesic ray $\gamma:[0,s^+)\to \mcM$ will be called
\emph{$s$--complete} if $s_+=\infty$ \emph{unless} there exists
some polynomial scalar invariant $\alpha$ such that
$$
 \limsup_{s\to s_+} |\alpha(\gamma(s))|=\infty
 \;.
$$
A similar definition applies to maximally extended geodesics
$\gamma:(s_-,s^+)\to \mcM$, with some polynomial scalar
invariant (not necessarily the same) unbounded in the
incomplete direction, if any. Here, by a \emph{polynomial
scalar invariant} we mean a scalar function which is a
polynomial in the metric, its inverse, the Riemann tensor and
its derivatives. It should be clear how to include in this
notion some other objects of interest, such as the norm
$g(X,X)$ of a Killing vector $X$, or of a Yano-Killing tensor,
etc. But care should be taken not to take scalars such as $\ln
(R_{ijkl} R^{ijkl})$ which could blow up even though the
geometry remains regular; this is why we restrict attention to
polynomials.

 A Lorentzian manifold $(\mcM ,g)$ will be said to be
\textit{$s$--complete} if every maximally extended geodesic is
$s$--complete. The notions of \textit{timelike
$s$--completeness}, or \textit{causal $s$--completeness} are
defined similarly, by specifying the causal type of the
geodesics in the definition above.

We have the following version of~\cite[Theorem~6.3,
p.~255]{kobayashi:nomizu} (compare also the \emph{Remark} on
p.~256 there), where geodesic completeness is weakened to
\emph{timelike $s$--completeness}:

\begin{Theorem}
 \label{Tscompl}
\textit{Let $(\mcM ,g)$, $(\mcM ',g')$ be analytic Lorentzian
manifolds of dimension $n+1$, $n\ge 1$, with
  $\mcM $ connected and simply connected, and $\mcM '$
  timelike $s$--complete. Then every isometric immersion $f_U:U \subset \mcM  \hookrightarrow
  \mcM '$, where $U$ is an open subset of $\mcM $,   extends
  uniquely to an isometric immersion $f:\mcM  \hookrightarrow \mcM '$.}
\end{Theorem}
\proof We need some preliminary lemmas which are proved as in
\cite{kobayashi:nomizu}, by replacing ``\textit{affine
mappings}" there by ``\textit{isometric immersions}":
\begin{Lemma}{\rm\cite[Lemma~1, p.~252]{kobayashi:nomizu}}
\label{a1}  {Let $\mcM $, $\mcM '$ be analytic manifolds, with
$\mcM $ connected. Let $f$, $g$ be analytic mappings $\mcM
\rightarrow \mcM '$. If $f$ and $g$ coincide on a nonempty open
subset of $\mcM $, then they coincide everywhere.}
\end{Lemma}

\begin{Lemma}{\rm\cite[Lemma~4, p.~254]{kobayashi:nomizu}}
\label{a3}
 {Let $(\mcM ,g)$ and $(\mcM ',g')$ be pseudo-Riemannian
manifolds of same dimension, with $\mcM $ connected, and let
$f$ and $g$ be isometric immersions of $\mcM $ into $\mcM '$.
If there exists some point $x \in \mcM $ such that $f(x)=g(x)$
and $f_*(X)=g_*(X)$ for every vector $X$ of $T_x\mcM $, then
$f=g$ on $\mcM $.}
%
\end{Lemma}

We can turn our attention now to the proof of
Theorem~\ref{Tscompl}. Similarly to the proof of Theorem 6.1 in
\cite{kobayashi:nomizu}, we define an \textit{analytic
continuation of $f_U$ along a continuous path}
$c:[0,1]\rightarrow \mcM $ to be a set of mappings $f_s$,
$0\leq s\leq 1$, together with a family of open subsets $U_s$,
$0\leq s\leq 1$, satisfying the properties:
\begin{itemize}
  \item $f_0=f_U$ on $U_0=U$;
  \item for every $s \in [0,1]$, $U_s$ is a neighborhood of
      the point $c(s)$ of the path $c$, and $f_s$ is an
      isometric immersion $f_s: U_s \subset \mcM
      \hookrightarrow \mcM '$;
  \item for every $s \in [0,1]$, there exists a number
      $\delta_s >0$ such that for all $s' \in [0,1]$,
      ($|s'-s|<\delta_s$)
   $\Rightarrow$ ($c(s')\in U_s \text{ and } f_{s'}=f_s$ in
   a neighborhood of $c(s')$);
\end{itemize}
We need to prove that, under the hypothesis of
$s$--completeness, such an analytic continuation does exist
along any  curve $c$. The argument is simplest for timelike
curves, so let us first assume that $c$ is timelike.
To do so, we consider the set:
\be A:=\{s \in [0,1]\ |\ \text{an analytic continuation exists
along $c$ on } [0,s]\} \ee
$A$ is nonempty, as it contains a neighborhood of $0$. Hence
$\bar{s}:= \sup A$ exists and is positive. We need to show that
in fact, $\bar{s}=1$ and can be reached.
Assume that this is not the case. Let $W$ be a normal convex
neighborhood of $c(\bar{s})$ such that every point $x$ in $W$
has a normal neighborhood containing $W$. (Such a $W$ exists
from Theorem 8.7, chapter III of \cite{kobayashi:nomizu}.) We
can choose $s_1<\bar{s}$ such that $c(s_1) \in W$, and we let
$V$ be a normal neighborhood of $c(s_1)$ containing $W$. Since
$s_1 \in A$, $f_{s_1}$ is well defined, and is an isometric
immersion of a neighborhood of $c(s_1)$ into $\mcM '$; we will
extend it to $V\cap I^\pm(c(s_1))$. To do so, we know that
$\exp:V^*\rightarrow V$ is a diffeomorphism, where $V^*$ is a
neighborhood of $0$ in $T_{c(s_1)}\mcM $, hence, in particular,
for $y \in V\cap I^\pm(c(s_1))$, there exists a unique $X \in
V^*$ such that $y=\exp X$. Define $X':= f_{s_{1}\ast} X$. Then
$X'$ is a vector tangent to $\mcM '$ at the point $f_{s_1}
(c(s_1))$.
Since $y$ is in the timelike cone of $c(\bar{s})$, $X$ is
timelike, and so is $X'$, as $f_{s_1}$ is isometric. We now
need to prove the following:
\begin{Lemma}
\textit{The geodesic $s\mapsto \exp(sX')$ of $\mcM '$ is well
defined for $0 \leq s \leq 1$.}
\end{Lemma}
 \proof  Let \be s^*:=\sup\{s
\in [0,1]\ |\ \exp(s'X') \text{ exists } \forall s' \in [0,s]
\}. \ee
First, such a $s^*$ exists, is positive, and we notice that if
$s^*<1$, then it is not reached. We wish to show that $s^*=1$
and is reached. Hence, it suffices to show that ``$s^*$ is not
reached" leads to a contradiction. Indeed, in such a case the
timelike geodesic $s \mapsto \exp(sX')$ ends at finite affine
parameter, thus, there exists a scalar invariant $\varphi$ such
that $\varphi(\exp(sX'))$ is unbounded as $s \rightarrow s^*$.
Now, for all $s<s^*$, we can define $h(\exp(sX)):=\exp(sX')$,
and this gives an extension $h$ of $f_{s_1}$ which is analytic
(since it commutes with the exponential maps, which are
analytic). By Lemma \ref{a3}, $h$ is in fact an isometric
immersion. By definition of scalar invariants we have
$$
 \varphi(\exp(sX'))=\tilde{\varphi}(\exp(sX))
 \;,
$$
where $\tilde{\varphi} $ is the invariant in $(\mcM ,g)$
corresponding to $\varphi$. But this is not possible since
$\tilde{\varphi}(\exp(sX))$ has a finite limit when
$s\rightarrow s^*$, and provides the desired contradiction.
\qed

\medskip

>From the last lemma we deduce that there exists a unique
element, say $h(y)$, in a normal neighborhood of
$f_{s_1}(c(s_1))$ in $\mcM '$ such that $h(y)=\exp(X')$. Hence,
we have extended $f_{s_1}$ to a map $h$ defined on $V\cap
I^\pm(c(s_1))$. In fact, $h$ is also an isometric immersion, by
the same argument as above, since it commutes with the
exponential maps of $\mcM $ and $\mcM '$. Then, since the curve
$c$ is timelike, this is sufficient to conclude that we can do
the analytic continuation beyond $c(\bar{s})$, since $V\cap
I^\pm(c(s_1))$ is an open set, and thus contains a segment of
the geodesic $c(s)$, for $s$ in a neighborhood of $\bar{s}$.

\begin{figure}[ht]
\begin{center}
\hspace{-2cm}{
 \psfrag{W}{\huge  $\!\!W$}
 \psfrag{cdeun}{\huge$ \!\!c(  1)$}
 \psfrag{csun}{\huge$ c(  s_1)$}
 \psfrag{cstilde}{\huge$ c(\tilde s)$}
 \psfrag{Ipcs}{\huge $\!I^+\big(c(\tilde s)\big)$}
 \psfrag{Imz}{\huge $\!\!\!\!\!I^- (z)$}
 \psfrag{zet}{\huge $z$}
\hspace{2.5cm}\resizebox{3in}{!}{\includegraphics{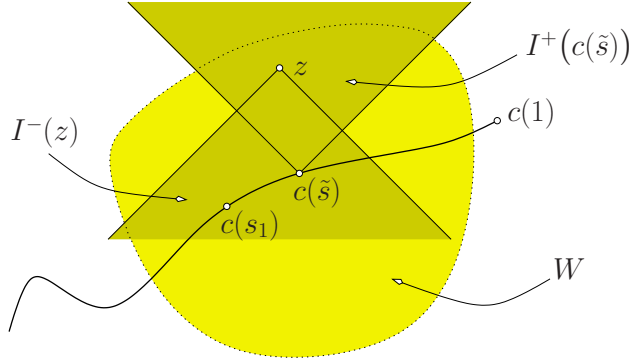}}
}
\caption{The analytic continuation at $c(\tilde s)$.
\label{FKobayashi}}
\end{center}
\end{figure}

Let us consider now a general, not necessarily timelike,
continuous curve  $c(s), 0\leq s \leq 1$, with $c(0) \in U$. As
before, we consider the set:
\be \{s \in [0,1]\ |\ \text{there exists an analytic
continuation of $f_U$ along } c(s'), 0\leq s'\leq s\}, \ee
and its supremum $\tilde{s}$. Assume that $\tilde{s}$ is not
reached. Let again $W$ be a normal neighborhood of
$c(\tilde{s})$ such that every point of $W$ contains a normal
neighborhood which contains $W$. Then, let $z$ be an element of
the set $I^+(c(\tilde{s}))\cap W$. $I^-(z)\cap W$ is therefore
an open set in $W$ containing $c(\tilde{s})$. Hence we can
choose $s_1 <\tilde{s}$ such that the curve segment
$c([s_1,\tilde{s}])$  is included in $I^-(z)\cap W$, see
Figure~\ref{FKobayashi}.
In particular, $z \in I^+(c(s_1))\cap W$. Since there exists an
analytic continuation up to $c(s_1)$, we have an isometric
immersion $f_{s_1}$ defined on a neighborhood $U_{s_1}$ of
$c(s_1)$, which can be assumed to be included in $W$. Hence,
from what has been seen previously, $f_{s_1}$ can be extended
as an isometric immersion, $\psi_1$, on $U_z:=U_{s_1}\cup
\left(I^+(c(s_1))\cap W\right)$, which contains $z$.
We now do the same operation for $\psi_1$ on $U_z$: we can
extend it by analytic continuation to an isometric immersion
$\psi_2$ defined on $U_z\cup\left(I^-(z)\cap W\right)$, which
is an open set containing the entire segment of the curve $x$
between $c(s_1)$ and $c(\tilde{s})$. In particular, $\psi_1$
and $\psi_2$ coincide on $U_z$, i.e. on their common domain of
definition; thus we obtain an analytic continuation of
$f_{s_1}$ along the curve $c(s)$, for $s_1\leq s\leq
\tilde{s}$; this continuation also coincides with the
continuation $f_s,\ s\in [s_1,\tilde{s}[\ $. This is in
contradiction with the assumption that $\tilde{s}$ is not
reached by any analytic continuation from $f_U$ along $x$.
Hence $\tilde{s}=1$ and is reached, that is to say we have
proved the existence of an analytic continuation of $f_U$ along
all the curve $x$.

The remaining arguments are as in~\cite{kobayashi:nomizu}.
\qed

\section{Geodesics}
 \label{sGeod}

We continue with a study of geodesics in $(\hmcM,\mhatg)$. Our
aim is to prove:

\begin{Theorem}
 \label{Tglobcontge}
All maximally extended causal geodesics in $(\hmcM,\mhatg)$ are
either complete, or reach a singular boundary $\{z=\xi_F\}$ in
finite affine time.
\end{Theorem}

The proof of Theorem~\ref{Tglobcontge} will occupy the
remainder of this section. We will analyze separately the
behavior of the geodesics in various regions of interest, using
coordinates suited for  the region at hand.

\medskip

\subsection{Geodesics in
the domain of outer communications away from the horizon}
 \label{sGdoc}

Whether in $\mcM_I$ or in $\hmcM$, the \emph{domain of outer
communications$^{\mbox{\scriptsize \rm\ref{Ffterm}}}$ $\doc$}
coincides with the set
$$
 \{z\in(\xi_3,\infty]\cup[-\infty,\xi_1]
 \}
 \;.
$$
We continue by showing that all geodesic segments  in $\doc$ of
finite affine length which do not approach the boundary
$\{z=\xi_3\}$ remain within compact sets of $\mcM$, with
uniform bounds on the velocity vector. This holds regardless of
the causal nature of the geodesic. To see this, let $s \mapsto
\gamma(s)$ be an affinely parameterized geodesic,
$$\gamma(s)=(t(s),\psi(s),z(s),x(s),\varphi(s))\; .$$
We have four constants of motion,
\be \lambda:=g(\dot{\gamma},\dot{\gamma})
 \;, \quad
 c_t := g(\partial_t,\dot{\gamma})
 \;, \quad
  c_\psi := g(\partial_\psi,\dot{\gamma})
 \;, \quad c_\varphi := g(\partial_\varphi,\dot{\gamma})\; .\ee
Written out in detail, keeping in mind that
$\sigma=\frac{1}{A}\sqrt{\frac{\nu}{\xi_F}}$,
\begin{eqnarray}
\label{energy}
& \lambda = -\frac{F(x)}{F(z)}\left(\dot{t}+\sigma(\xi_{1}-z)\dot{\psi}\right)^{2}
   +\frac{F(z)}{A^{2}(x-z)^{2}}\left[-F(x)\left(\frac{\dot{z}^{2}}{G(z)} +\frac{G(z)}{F(z)}\dot{\psi}^{2}\right)\right.
\\
  & \hspace{7cm}\left. +F(z)\left(\frac{\dot{x}^{2}}{G(x)}+\frac{G(x)}{F(x)}\dot{\varphi}^{2}\right)\right]\;\nonumber
   ;
\end{eqnarray}
\bel{ct}
c_t=-\frac{F(x)}{F(z)}\left(\dot{t}+\sigma(\xi_1-z)\dot{\psi}\right)\;
; \ee
\bel{cpsi}
c_\psi=\sigma(\xi_1-z)c_t-\frac{G(z)F(x)}{A^{2}(x-z)^{2}}\dot{\psi}\;
;\ee
\bel{cphi}
c_\varphi=\frac{F^2(z)G(x)}{A^{2}(x-z)^{2}F(x)}\dot{\varphi}\;
. \ee
This leads to
\bel{dotpsi}
\dot{\psi}=\frac{A^2(x-z)^2}{F(x)G(z)}\left(\sigma(\xi_1-z)c_t-c_\psi\right)\;
,\ee
\bel{dott}
\dot{t}=-\frac{F(z)}{F(x)}c_t-\sigma(\xi_1-z)\frac{A^2(x-z)^2}{F(x)G(z)}\left(\sigma(\xi_1-z)c_t-c_\psi\right)\;
,\ee
\bel{dotphi}
G(x)\dot{\varphi}=\frac{A^2(x-z)^2F(x)}{F(z)^2}c_\varphi\; ,\ee
\begin{eqnarray}
\label{lambda}
& \lambda=-\frac{F(z)}{F(x)}c_t^{2}
   -\frac{F(z)F(x)}{A^{2}(x-z)^{2}G(z)}\dot{z}^{2}-\frac{A^2(x-z)^2}{F(x)G(z)}\left(\sigma(\xi_1-z)c_t-c_\psi\right)^{2}
\\
  & \hspace{5cm} +\frac{F^2(z)}{A^2(x-z)^2}\frac{\dot{x}^{2}}{G(x)}
   +\frac{A^2(x-z)^2F(x)}{G(x)F^2(z)}c_\varphi ^2 \; .\nonumber
\end{eqnarray}
We have:
\begin{enumerate}
 \item Those geodesics for which $\lim\inf_{s\to\infty}
     (x(s)-z(s))=0$ can be studied by transforming the
     metric to explicitly asymptotically flat coordinates
     as in~\cite{EmparanReall}, and using known methods
     (see, e.g.,~\cite[Appendix~B]{ChmassCMP}
     and~\cite[Appendix]{poorman2}). Such geodesics
     eventually remain in the asymptotically flat region
     and are complete in the relevant direction. So,
     without loss of generality we can assume in the
     remainder of our analysis that
     \bel{epsin}
     |x-z|\ge \epsilon_0
     \ee
     for some $0<\epsilon_0<1$.
\item Consider those geodesic segments for which
    $$
    2 \xi_1\le z(s) \le \xi_1
    \;.
    $$
In this region the functions $z$ and $x$ are related to
polar-type coordinates near axes of rotation $G(z)=0$ and
$G(x)=0$; in fact, well behaved polar-type coordinates
$(\myrho ,\mu)$ are obtained by introducing
\bel{rhomudef} d\myrho  = \frac{ dx}{\sqrt{G(x)}}\;,\quad
d\mu =\frac{ dz}{\sqrt{|G(z)|}}  \;. \ee
We then rewrite \eq{lambda} as
\begin{eqnarray}
\label{lambda2}
   & F(x) {\dot{\mu }^{2} }
+{F(z)} {\dmyrho ^{2}}
 +\frac{A^4(x-z)^4F(x)}{G(x)F^3(z)}c_\varphi ^2 &
 \\
& {+\frac{A^4(x-z)^4}{F(x)F(z)|G(z)|}\left(\sigma(\xi_1-z)c_t-c_\psi\right)^{2}
} &= \ \ \frac{A^{2}(x-z)^{2}}{F(z)}\Bigg[\lambda
+\frac{ F(z)}{F(x)}c_t^{2} \Bigg] \; .
\nonumber
\end{eqnarray}
The right-hand-side is bounded by a constant $C$, while the
coefficients $F(x)$ of $\dot \mu^2$ and $F(y)$ of
$\dot\myrho ^2$ are bounded from above and away from zero,
so there exists a constant $C_1$ such that
\begin{eqnarray}
  {\dot{\mu }^{2} }
+ {\dmyrho ^{2}} \le C_1
\;.
\label{lambda2a}
\end{eqnarray}
Inspecting \eq{dotpsi}-\eq{dotphi}, and noting that the
zero of $G(z)$ in the denominator of the right-hand-side of
\eq{dott} is canceled by the $z-\xi_1$ factor in the
numerator, we find that there exists a constant $C_2$ such
that
\begin{eqnarray}
  {\dot{t }^{2} }
+ {\dmyrho ^{2}} + {\dot{\mu }^{2} }
+ G^2(z){\dot{\psi}^{2}}+ G^2(x){\dot{\varphi}^{2} } \le C_2
\;.
\label{lambda2b}
\end{eqnarray}
It follows from \eq{lambda2} that a non-zero $c_\varphi$ prevents
$x$ from approaching $\xi_1$ and $\xi_2$ unless $x-z\to 0$,
similarly a non-zero $c_\psi$ prevents $z$ from approaching $\xi_1$
unless $x-z\to 0$. So, under \eq{epsin}, we find a bound on $|\dot
\psi|$  from \eq{dotpsi} when $c_\psi$ is zero (since then a factor
$z-\xi_1$ in $G(z)$
is cancelled by a similar factor in the numerator), or from
\eq{lambda2b} otherwise. A similar analysis of $\dot \varphi$ allows
us to conclude that
\begin{eqnarray}
  {\dot{t }^{2} }
+ {\dmyrho ^{2}} + {\dot{\mu }^{2} }
+  {\dot{\psi}^{2}}+  {\dot{\varphi}^{2} } \le C_3
\;.
\label{lambda2c}
\end{eqnarray}

\item Consider, next,  geodesic segments for which
    $$
    -\infty\le z \le 2 \xi_1 \ \mbox{ or } \ \xi_3+\epsilon \le z \le \infty
    \;,
    $$
    where $\epsilon$ is some strictly positive number.
    Introducing $Y=-1/z$, from \eq{lambda} we find
\begin{eqnarray}
\label{lambda3b}
 \lefteqn{
   \frac{|F(z)|F(x)}{A^{2}(x-z)^{2}|G(z)|Y^4}\dot{Y}^{2}
+\frac{F^2(z)}{A^2(x-z)^2} {\dmyrho ^{2}
   +\frac{A^2(x-z)^2F(x)}{G(x)F^2(z)}c_\varphi ^2}}
 &&
 \\
&&=\lambda
+\frac{ F(z)}{F(x)}c_t^{2}
+\frac{A^2(x-z)^2}{F(x)G(z)}\left(\sigma(\xi_1-z)c_t-c_\psi\right)^{2} \; .\nonumber
\end{eqnarray}
By an argument similar to the above, but simpler, we obtain
\begin{eqnarray}
  {\dot{t }^{2} }
+ {\dmyrho ^{2}} + {\dot{Y}^{2} }
+  {\dot{\psi}^{2}}+  {\dot{\varphi}^{2} } \le C_4
\;.
\label{lambda3c}
\end{eqnarray}
Here one has to use a cancelation in the coefficient of
$c_t^2$ in \eq{lambda3b}, as well as in the coefficient of
$c_t$ in   \eq{dott}, keeping in mind that
$\sigma=\frac{1}{A}\sqrt{\frac{\nu}{\xi_{F}}}$; e.g.,
 \bel{dott2} \dot{t}=-\frac 1
 {F(x)}\underbrace{\Bigg(\underbrace{{F(z)}}_{= 1/(Y\xi_F)
 +O(1)}
 +\underbrace{\sigma^2(\xi_1-z)^2\frac{A^2(x-z)^2}{G(z)}}_{=-\sigma^2
 A^2/(\nu Y)+O(1)} \Bigg)}_{=O(1)}c_t +O(1)\; ,
 \ee
where $O(1)$ denotes terms which are bounded as $Y\to 0$.
\end{enumerate}
Usual considerations about maximally extended geodesics show
now that:

\begin{Proposition}
 \label{Pfirstregion}
For any $\epsilon>0$,   {the geodesics which are maximally
extended within
the region $z\in[\xi_3+\epsilon,\infty]\cup[-\infty,\xi_1]$ are
either complete, or acquire a smooth end point at
$\{z=\xi_3+\epsilon\}$.}
%
\end{Proposition}

\medskip

\subsection{Geodesics in the region $\{\xi_F<z<\xi_3\}$}
 \label{S5.2}

In this coordinate range both   $F(z)$   and $G(z)$ are
negative, and we rewrite \eq{lambda} as
\begin{eqnarray}
  \frac{|F(z)|}{|G(z)|} {\dot{z }^{2} }
-\frac{F^2(z)}{F(x)} {\dmyrho ^{2}}
 = \mychitwo+
\frac{A^4(x-z)^4 }{G(x)F^2(z)}c_\varphi ^2
\;,
\label{lambda2xx}
\end{eqnarray}
where we have set
\begin{eqnarray}
\label{lambda2xx2}
& \phantom{xxx} \mychitwo:=
  \frac{A^4(x-z)^4}{F^2(x) (-G(z))}\left(\sigma(\xi_1-z)c_t-c_\psi\right)^{2}
 +\frac{A^{2}(x-z)^{2}}{F(x)}\Bigg[-\lambda
+\frac{| F(z)|}{F(x)}c_t^{2} \Bigg] \; .
\end{eqnarray}

\medskip

\subsubsection{Timelike geodesic incompleteness}
 \label{sstgi}
The extended space-time will not be geodesically complete if
one can find a maximally extended geodesic with finite affine
length. Consider, thus any future directed, affinely
parameterized timelike geodesic $\gamma$ entirely contained in
the region $\{\xi_F<z<\xi_3\}\cap \{\hat v>0, \hat w>0\}$,
 and maximally
extended there; an identical argument applies to past directed
timelike geodesics in the region $\{\xi_F<z<\xi_3\}\cap  \{\hat
v<0,\hat w<0\}$. Since $z$ is a time function in this region,
$z$ is strictly decreasing along $\gamma$. {}From
\eq{lambda2xx} we have
$$
   \frac{F(z)F(x)}{A^{2}(x-z)^{2}G(z)}\dot{z}^{2}\ge -\lambda
   \;,
$$
which gives $\sqrt{|F(z)|}|\dot z|\ge \epsilon
\sqrt{|\lambda|}\sqrt{|G(z)|}>0$ for some constant $\epsilon$.
The proper time parameterization  is obtained by choosing
$\lambda=-1$. Let $L(\gamma)$ denote the proper length along
$\gamma$; keeping in mind that $\dot z=dz/ds$ we obtain
$$
 L(\gamma)
 =
 \int_{\xi_F}^{\xi_3} \left|\frac {ds}{dz}\right| dz\le \frac 1
 \epsilon \int_{\xi_F}^{\xi_3} \sqrt{\frac {{F(z)}} {G(z)}} dz<\infty
 \;.
$$
Hence every such geodesic reaches the singular boundary
$\{z=\xi_F\}$ in finite proper time  \emph{unless} $(\dot
z,\dot \theta)$ becomes unbounded before reaching that set. We
will see shortly that this second possibility cannot occur.

\medskip

\subsubsection{Uniform bounds}
 \label{ssUb}
We wish, now, to derive uniform bounds on  timelike geodesic
segments contained in the region $\{\xi_F+\epsilon<z<\xi_3\}$,
with any $\epsilon>0$.

We start by noting that
\bel{dotpsi3} \mydot{\hat
\psi}=\frac{A^2(x-z)^2(\xi_3-z)}{F(x)G(z)(\xi_3-\xi_1)}\left(\sigma(\xi_1-z)c_t-c_\psi\right)
+\frac{F(z)}{\mygamma (\xi_3-\xi_1)F(x)}c_t
 \;.
\ee
which is well behaved throughout the region of current
interest.

In the region $\{\hat v >0\;, \ \hat w>0\}$ we can introduce
coordinates $v$ and $w$ using the formulae
$$
 v = \frac {\ln \hat v } c\;,\quad  w = -\frac { \ln \hat w} c
 \;,
$$
and then define $t$ and $z$ using \eq{12.4}-\eq{12.5}.  With
those definitions one recovers the form \eq{metric20} of the
metric, so that we can use the previous formulae for geodesics:
\begin{equation}
\label{dv1}
  \mydot {\hat v}= c \Bigg\{-\hat v\frac{F(z)}{F(x)}c_t
  -\frac{\sigma }{\nu \hat w(z-\xi_2)^2}
  \Bigg[\beta(x,z)
 + {\sqrt{-F(\xi_3)}} \mydot  z \Bigg]\Bigg\}\;,
\end{equation}
\begin{equation}
 \label{dw1}
 \mydot {\hat w} = c \Bigg\{\frac{F(z)}{F(x)}
 c_t\hat w +\frac{\sigma }{\nu \hat v(z-\xi_2)^2}\left[\beta(x,z)
 - {\sqrt{-F(\xi_3)}} \mydot  z \right]\Bigg\}\;,
\end{equation}
where
$$
\beta(x,z):=\frac{A^2(x-z)^2}{F(x)}\left(\sigma(\xi_1-z)c_t-c_\psi\right)\;,
$$
and we note that both right-hand-sides have a potential problem at
$\{z=\xi_3\}$, where $\hat v \hat w$ vanishes. Next, from \eq{dv1}
and \eq{dw1},
\bel{alphadef2}
 \underbrace{\sqrt{-F(z)}\mydot z}_{=:\alpha} =
-\sqrt{\left|\frac{F(z)}{F(\xi_3)}\right|} \frac {\nu
(z-\xi_2)^2}{2c\mygamma
 }\left( \hat w \mydot {\hat v} + \hat v \mydot {\hat w}\right)
 \;,
\ee
while from \eq{dv1}-\eq{dw1} we further have
\bel{dott5}
 \hat w \mydot {\hat v} - \hat v \mydot {\hat w}=
 -\frac{2cF(z)}{F(x)}c_t\hat v \hat
 w-\frac{2c\sigma }{\nu (\xi_2-z)^2}\frac{A^2(x-z)^2}{ F(x)}
\left(\sigma(\xi_1-z)c_t-c_\psi\right)   \;
 .
\ee
We continue by rewriting \eq{lambda} so that the problematic factors
in \eq{dv1}-\eq{dw1} are grouped together
\begin{equation}
\label{lambda5}
    \frac {F(x)}{|G(z)|{A^{2}(x-z)^{2}}}
   \Bigg[ \underbrace{|F(z)|\bmydot {z}^{2}
   -\frac{A^4(x-z)^4}{F(x)^2}\left(\sigma(\xi_1-z)c_t-c_\psi\right)^{2}}_{\alpha^2-\beta^2=(\alpha-\beta)(\alpha+\beta)}\Bigg]
\end{equation}
\begin{equation}
  = -\lambda+\frac{| F(z)|}{F(x)}c_t^{2}+\frac{F^2(z)}{A^2(x-z)^2}{\bmydot {\myrho }^{2}}
   +\frac{A^2(x-z)^2F(x)}{G(x)F^2(z)}c_\varphi ^2 \; .
   \nonumber
\end{equation}

By Section~\ref{sstgi} any causal geodesic will either reach
$\{z=\xi_3\}$ in finite affine time, say $s=\mathring s$, or
will cease to exist before that time. In what follows we
therefore assume $0\le s < \mathring s$.

We continue by writing down the evolution equations for $x$ and
$z$, which can easily be obtained from the Lagrangean $L=g(\dot
\gamma, \dot \gamma)$, and read
\begin{eqnarray}
\label{xeq}
&&
\\
& \hspace{-9cm} 2 \frac{d}{ds}\left(\frac{F^2(z) }{A^2 (x-z)^2
G(x)}\frac {dx}{ds}\right) &  \nonumber
\\
 & = -\frac{F'(x)F(z) }{ F^2(x)}c_t^2  - \frac 1 {A^2 G(z)}
\frac
\partial {\partial x}\left( \frac {F(x)}{(x-z)^2}\right)\left( F(z)\dot
z^2 + \frac {A^4 (x-z)^4}{F^2(x)}(\sigma
(\xi_1-z)c_t-c_\psi)^2\right)&
 \nonumber
 \\
 & \phantom{=}
 + \frac {F^2(z)}{A^2}\left[\frac
\partial {\partial x} \left( \frac 1
 {G(x)(x-z)^2}\right) \dot x^2 + \frac
\partial {\partial x} \left(
 \frac{G(x)}{F(x)(x-z)^2}\right) \frac {A^4 (x-z)^4
 F^2(x)}{F^4(z)G^2(x)}c_\varphi^2\right]
 \;.&
 \nonumber
 \end{eqnarray}
\begin{eqnarray}
\label{zeq}
&&
\\
 & \hspace{-7cm} -2 \frac{d}{ds}\left(\frac{F(x)F(z) }{A^2
(x-z)^2  G(z)}\frac {dz}{ds}\right)=  \frac{F'(z)  }{ F (x)}c_t^2   & \nonumber
\\
&  - 2\sigma \frac{A^2 (x-z)^2}{F(x)G(z)}(\sigma(\xi_1 - z)c_t - c_\psi)c_t +\frac
\partial {\partial z} \left( \frac
{F^2(z)}
 {A^2(x-z)^2}\right) \left[  \dot \myrho^2 +  \frac {A^4 (x-z)^4
 F (x)}{F^4(z)G (x)}c_\varphi^2\right]&
 \nonumber
 \\
 &
 \phantom{=}
 - \frac {F(x)} {A^2 }\left[
\frac
\partial {\partial z} \left( \frac {F(z)}{G(z)(x-z)^2}\right) \dot z^2 +
\frac
\partial {\partial z} \left( \frac {G(z)}{(x-z)^2}\right) \frac{A^4
(x-z)^4}{F^2(x)G^2(z)}(\sigma (\xi_1-z)c_t-c_\psi)^2\right]&
 \;.
 \nonumber
\end{eqnarray}

Since $z$ is a time function, the derivative $\dot z$ has
constant sign. So $z$ can be used as a parameter along
$\gamma$, and we can view \eq{zeq} as an evolution equation in
$z$ for $\dot z$. For this we multiply by $ds/dz$, obtaining
\begin{eqnarray}
\label{zeq2}
&&
\\
 & \hspace{-12cm}\lefteqn{-2 \frac{d}{dz}\left(\frac{F(x)F(z) }{A^2
(x-z)^2 G(z) }\frac {dz}{ds}\right)} & \nonumber
\\
& \hspace{-3cm}=  \left[\frac{F'(z)  }{ F (x)}c_t^2  - 2\sigma \frac{A^2
(x-z)^2}{F(x)G(z)}(\sigma(\xi_1 - z)c_t - c_\psi)c_t
 \right]
    \frac{ds}{dz}&
 \nonumber
 \\
 & \hspace{-3cm}
 \phantom{=}
 + \frac
\partial {\partial z} \left( \frac {F^2(z)}
 {A^2(x-z)^2}\right) \left[  \left(\frac {d\myrho }{dz}\right)^2 \frac {dz}{ds} +  \frac {A^4 (x-z)^4
 F (x)}{F^4(z)G (x)}c_\varphi^2\frac {ds}{dz}\right]&
 \nonumber
 \\
 &
 \phantom{=}
 - \frac {F(x)} {A^2 }\left[
\frac
\partial {\partial z} \left( \frac {F(z)}{G(z)(x-z)^2}\right)\frac
{dz}{ds} + \frac
\partial {\partial z}\left( \frac {G(z)}{(x-z)^2}\right)
\frac{A^4 (x-z)^4}{F^2(x)G^2(z)}(\sigma
(\xi_1-z)c_t-c_\psi)^2\frac {ds}{dz}\right]
 \;. &
 \nonumber
\end{eqnarray}

Yet another variation on \eq{lambda2xx} reads
\begin{eqnarray}
  \frac{|F(z)|}{|G(z)|}   =
\frac{F^2(z)}{F(x)} \left(\frac {d\myrho} {dz}\right)^2
 + \left(\frac {d s} {dz}\right)^2\left(
 \mychitwo
 +
\frac{A^4(x-z)^4 }{G(x)F^2(z)}c_\varphi ^2 \right) \; .%
\label{lambda2xxx}
\end{eqnarray}
To obtain uniform bounds as $z$ approaches $\xi_3$, one can
proceed as follows: Let $f>0$ be defined by the formula
\bel{fdef}
 f:= \frac{F(x)F(z) }{A^2
(x-z)^2 G(z) }\frac {dz}{ds}  \;. \ee
Using the identity
\bel{eldifid}
  -\frac 2 {\sqrt{-G(z)}} \frac d {dz}\left(\sqrt{-G(z)} f\right) = -2 \frac {df}{dz} - \frac{G'(z)}{G(z)} f
\ee
we rewrite \eq{zeq2} as
\begin{eqnarray}
\label{zeq2ag}
&&
\\
  & \hspace{-7cm} -\frac{2}{
\sqrt{-G(z)}}\frac{d}{dz}\left(\frac{F(x)|F(z)| }{A^2 (x-z)^2
\sqrt{-G(z)} }\frac {dz}{ds}\right) & \nonumber
\\
& \hspace{-4cm} =  \left[\frac{F'(z)  }{ F (x)}c_t^2 - 2\sigma \frac{A^2
(x-z)^2}{F(x)G(z)}(\sigma(\xi_1 - z)c_t - c_\psi)c_t
    \right]\frac {ds}{dz}&
 \nonumber
 \\
 & \hspace{-4cm}
 \phantom{=}
  + \frac
\partial {\partial z} \left( \frac {F^2(z)}
 {A^2(x-z)^2}\right) \left[  \left(\frac {d\myrho }{dz}\right)^2 \frac {dz}{ds} +  \frac {A^4 (x-z)^4
 F (x)}{F^4(z)G (x)}c_\varphi^2\frac {ds}{dz}\right]&
 \nonumber
 \\
 &
 \phantom{=}
 - \frac {F(x)} {A^2 G(z)}\left[
\frac
\partial {\partial z} \left( \frac {F(z)}{(x-z)^2}\right)\frac
{dz}{ds} + \frac
\partial {\partial z}\left( \frac {G(z)}{(x-z)^2}\right)
\frac{A^4 (x-z)^4}{F^2(x)G(z)}(\sigma
(\xi_1-z)c_t-c_\psi)^2\frac {ds}{dz}\right]
 \;.&
 \nonumber
\end{eqnarray}
\Eq{lambda2xxx} gives
\begin{eqnarray}
\\
\nonumber
  \hspace{-7cm} \frac{F(x)}{F^2(z)} \frac{d z}{ds}
 \Bigg[
\frac{F^2(z)}{F(x)} \left(\frac {d\myrho} {dz}\right)^2
 + \left(\frac {d s} {dz}\right)^2
 \frac{A^4(x-z)^4 }{G(x)F^2(z)}c_\varphi ^2\Bigg]
 \\
   =\frac{F(x)}{|F(z)G(z)|} \frac{d z}{ds}
-\mychitwo \frac{F(x)}{F^2(z)}\frac {d s} {dz}  \; , \nonumber %
 \label{lambda2xxxr}
\end{eqnarray}
leading to
\begin{eqnarray}
\label{zeq2ag3} &\hspace{-5.2cm}\frac{2}{
\sqrt{-G(z)}}\frac{d}{dz}\left(\frac{F(x)|F(z)| }{A^2 (x-z)^2
\sqrt{-G(z)} }\frac {dz}{ds}\right) & 
\\
& = \frac {F(x)} {A^2 G(z)} \Bigg\{ \underbrace{\frac
\partial {\partial z} \left( \frac {F(z)}{(x-z)^2}\right)
  - \frac
\partial {\partial z} \left( \frac {F^2(z)}
 {(x-z)^2}\right)\frac{1}{F(z)}
 }_{-F'(z)/(x-z)^2}
    \Bigg\} \frac {dz}{ds}%
-
 \frac {ds}{dz} \mychione
 \;,& \nonumber
\end{eqnarray}
where
\begin{eqnarray}
\label{MLIx1}
 \phantom{xxx}\mychione&:= &
 \frac{F'(z)  }{ F (x)}c_t^2    - 2\sigma \frac{A^2
(x-z)^2}{F(x)G(z)}(\sigma(\xi_1 - z)c_t - c_\psi)c_t
 \\
 &&
 \phantom{=}
-
  \frac
\partial {\partial z}\left( \frac {G(z)}{(x-z)^2}\right)
\frac{A^2 (x-z)^4}{F(x)G^2(z)}(\sigma (\xi_1-z)c_t-c_\psi)^2
 \nonumber
 \\
 &&
 \phantom{=}
 +  \frac
\partial {\partial z} \left( \frac {F^2(z)}
 {A^2(x-z)^2}\right)\Bigg[ {\frac{A^4(x-z)^4}{F (x) F^2(z) G(z) }\left(\sigma(\xi_1-z)c_t-c_\psi\right)^{2}}
  \nonumber
  \\
  &&
 \phantom{=}+
 \frac{A^{2}(x-z)^{2}}{F^2(z)}\Bigg(\lambda
+\frac{ F(z) }{F(x)}c_t^{2} \Bigg)\Bigg]
 \;.\nonumber
\end{eqnarray}
Using an identity similar to \eq{eldifid} with $G$ replaced by
$F$, this can also be rewritten as
\bea
  2\frac{1}
{\sqrt{|F(z)G(z)|}}\frac{d}{dz}\left(\frac{F(x)
(|F(z)|)^{3/2}}{A^2 (x-z)^2 \sqrt{ |G(z)} |}\frac
{dz}{ds}\right)= -  \eta \mychione
 \frac {ds}{dz}
  \;;
\eeal{zeq2ag4}
here, we have written the general formula which holds whatever
the sign of $F(z)G(z)$, with $\eta=\pm 1$ being that sign; in
the current context, $\eta=1$. Setting
$$
 h:= \frac{F(x)
(-F(z))^{3/2}}{A^2 (x-z)^2 \sqrt{|G(z)|} }\frac
{dz}{ds}
 \;,
$$
we obtain
\bel{zeq2ag5}
 2h\frac {dh}{dz}= \frac {dh^2}{dz} = -\frac{F(x)
 F^2(z)}{A^2(x-z)^2}  \eta \mychione
 \;.
\ee
>From \eq{lambda2xxx} one has
\begin{eqnarray}
 {
 \left|\frac {d\myrho} {dz}\right| \le
  \sqrt{\frac{F(x)}{|F(z)G(z)|}}
\;.  }
\label{lambda2xxxx}
\end{eqnarray}
Since the right-hand-side is integrable in $z$ on
$[\xi_F,\xi_3]$, we infer that $\myrho$ has finite limits both
as $z\to \xi_F$ and $z\to \xi_3$,
\bel{thetalimits}
 \theta\to_{z\to \xi_3} \theta_3\;, \quad x \to_{z\to \xi_3} x_3\;, \qquad
 \theta\to_{z\to \xi_F} \theta_F\;, \quad x \to_{z\to \xi_F} x_F\;.
\ee

In everything that follows we choose some small $\epsilon>0$
and assume that $z\in [\xi_F+\epsilon, \xi_3)$.

Suppose, first, that
\bel{restcond}
 \sigma(\xi_1-\xi_3)c_t-c_\psi=0 \ \Longrightarrow \ \sigma(\xi_1-z)c_t-c_\psi = \sigma(\xi_3-z)c_t
 \;.
\ee
Then the right-hand-side of \eq{zeq2ag5} is bounded, which
implies that $h$ is bounded, and has a limit as $z$ approaches
$\xi_3$. From the definition of $h$ we conclude that
$$
 |\dot z| \le C \sqrt{-G(z)}
 \;.
$$
\Eq{lambda2xx} implies, for $z$ near $\xi_3$,
\begin{eqnarray}
   {\dmyrho ^{2}}
 +
\frac{ c_\varphi ^2  }{G(x) }\le C
\;.
\label{MLI1}
\end{eqnarray}
We  rewrite \eq{dv1}-\eq{dw1} as  evolution equations in $z$:
\begin{equation}
\label{dv1x}
 \mydotz {\hat v}= c\hat v \Bigg\{-\frac{F(z)}{F(x)}c_t\mydotz s
  +
  \frac{\sigma (z-\xi_1)}{G(z)}
  \Bigg[\beta(x,z) \mydotz s
 + {\sqrt{-F(\xi_3)}} \Bigg]\Bigg\}\;,
\end{equation}
\begin{equation}
\label{dw1x} \mydotz {\hat w} = c \hat w \left\{\frac{F(z)}{F(x)}
 c_t \mydotz s
 -
 \frac{\sigma (z-\xi_1)}{G(z)}\left[\beta(x,z) \mydotz s
 - {\sqrt{-F(\xi_3)}}   \right]\right\}\;,
\end{equation}
where $\beta(x,z)$ has been defined as:
$$
\beta(x,z):=\frac{A^2(x-z)^2}{F(x)}\left(\sigma(\xi_1-z)c_t-c_\psi\right)\;.
$$
Note that \eq{restcond} implies that all prefactors of $\mydotz
s$ are bounded near $z=\xi_3$.

Suppose, first, that $c_\psi =0$, by \eq{restcond} this happens
if and only if $c_t=0$. Comparing \eq{dv1x} with \eq{dw1x} one
finds that
$$
 \frac {d(\ln \hat w)}{dz}=\frac {d(\ln \hat v)}{dz}
 \quad \Longleftrightarrow \quad
 \frac {d(\ln (\hat w/\hat v))}{dz} =0
\;.
$$
Thus  there exists a constant $\rho$, different from $0$ since
$\hat v \hat w \ne 0$ in the region of interest, such that
$$
 \hat v = \rho \hat w
 \;.
$$
Inserting back into \eq{dv1}  one finds
$$
\left|\mydot {\hat v}\right| \le  C \frac{|\dot z|}{\hat w}\le
 C\frac{\sqrt{\hat v \hat w}}{\hat w} =   C\frac{\sqrt{\hat v  }}{\sqrt{  \hat w}}
\le C
 \;,
$$
with a similar calculation for $d\hat w/ds$ using \eq{dw1}.
Thus
\bel{ML36}
 \left|\mydot {\hat v}\right| +
 \left|\mydot {\hat w}\right| \le C
 \;.
\ee
>From \eq{lambda2xxx}, $\dot \theta$ is bounded, as well as
derivatives of all coordinates functions along $\gamma$, and
smooth extendibility of the geodesic across $\hat w=\hat v=0$
readily follows.

So, still assuming \eq{restcond}, we suppose instead that $c_t$
is different from zero.  From \eq{lambda2xxx}, and since
$\mychitwo $ is positive, we get
\bel{ML34}
 \left|\mydotz s\right| \le \frac C {\sqrt{|G(z)|}}
 \;.
\ee
 \Eq{dv1x} with \eq{dw1x} gives now
$$
 \left|
 \frac {d(\ln (\hat w/\hat v))}{dz} \right| \le \frac C {\sqrt{|G(z)|}}
\;,
$$
which is integrable in $z$, so
$$
 \hat v = \rho (z) \hat w
 \;,
$$
for a function $\rho$ which has a finite limit as $z\to \xi_3$.
One concludes as when $c_t=0$.

We continue with the general case,
\bel{ML48}
 \sigma(\xi_1-\xi_3)c_t-c_\psi\ne 0
 \;.
\ee
Near $\xi_3$, the $x$--dependence of the most singular terms in
\eq{zeq2ag5} cancels out, leading to
\beal{ML41}
 \phantom{xx}
 \frac{dh^2}{dz}
  & = &
  \frac{ F^2(\xi_3)\Big(\sigma(\xi_1-\xi_3)c_t
- c_\psi\Big)^2+O(z-\xi_3)}{\nu
 (\xi_3-\xi_2)(\xi_3-\xi_1)}\times \frac{1}{(z-\xi_3)^2}
\\
  &=: &\frac{a^2+O(z-\xi_3)}{(z-\xi_3)^2}
 \;.
  \nonumber
\eea
By integration,
\bel{ML41a}
 h^2=-\frac{a^2 }{z-\xi_3} + O ( \ln |z-\xi_3|)
 \;.
\ee
Then, from the definition of $h$, we eventually find
\bel{ML49}
 \frac{dz}{ds}=
\underbrace{\frac{
A^2(x_3-\xi_3)^2}{\sqrt{|F(\xi_3)|}F(x_3)}(\sigma(\xi_1-\xi_3)c_t
- c_\psi)}_{=:\mathring a} +O ( (z-\xi_3)\ln |z-\xi_3|)
 \;.
\ee

Next, we need to know how the $x$--limit is attained. For this,
integration of \eq{lambda2xxxx} gives
$$
 |\theta(z)-\theta(\xi_3)| \le C\sqrt{|z-\xi_3|}
 \;.
$$
But, by the definition  \eq{rhomudef} of $\theta$, we have, for
$\theta_1$ close to $\theta_2$,
$$
|\theta_1-\theta_2|\approx \left\{
                             \begin{array}{ll}
                               \sqrt{ |x_1-x_2|}, & \hbox{$G(x_2)=0$;} \\
                               |x_1-x_2|, & \hbox{otherwise.}
                             \end{array}
                           \right.
$$
Hence
$$
 |x(z)-x(\xi_3)| \le  \left\{
                             \begin{array}{ll}
                               C |z-\xi_3|, & \hbox{$G(x_3)=0$;} \\
                              C \sqrt{|z-\xi_3|}, & \hbox{otherwise.}
                             \end{array}
                           \right.
 \;.
$$
Inserting this into \eq{lambda2xx}, the leading order
singularity cancels out:
\begin{eqnarray}
   {\dmyrho ^{2}}
 +
\frac{ c_\varphi ^2  }{G(x) }\le C
   |z-\xi_3|^{-1/2}
\;.
\label{ML44}
\end{eqnarray}

We pass now to the equation satisfied by $\dot x$. In terms of
$\myrho$, \eq{xeq} can be rewritten as
\begin{eqnarray}
\label{xeq3}
&&
\\
 &\hspace{-9cm} \frac 2{\sqrt {G(x)}}
\frac{d}{ds}\left(\frac{F^2(z) }{A^2 (x-z)^2 }\frac {d\myrho
}{ds}\right) & \nonumber
\\
& = -\frac{F'(x)F(z) }{ F^2(x)}c_t^2 - \frac 1 {A^2 G(z)}
\frac
\partial {\partial x}\left( \frac {F(x)}{(x-z)^2}\right)\left( F(z)\dot
z^2 + \frac {A^4 (x-z)^4}{F^2(x)}(\sigma
(\xi_1-z)c_t-c_\psi)^2\right)&
 \nonumber
 \\
 &
 \phantom{=}
 + \frac {F^2(z)}{A^2}\left[\frac
\partial {\partial x} \left( \frac {1}
 { (x-z)^2}\right)  \dot \myrho^2 + \frac
\partial {\partial x} \left(
 \frac{G(x)}{F(x)(x-z)^2}\right) \frac {A^4 (x-z)^4
 F^2(x)}{F^4(z)G^2(x)}c_\varphi^2\right]
 \;. & \nonumber
\end{eqnarray}
We can use \eq{lambda2xx} to eliminate $\dot z^2$ from this
equation, obtaining
\bea
 {\frac 2{\sqrt {G(x)}}
 \frac{d}{ds}\left(\frac{F^2(z) }{A^2 (x-z)^2  }\frac {d\myrho
 }{ds}\right)}  = \mychithree
 -    \frac {F'(x)F^2(z)}{A^2F(x)(x-z)^2}
{\dmyrho ^{2}}
 \;,
\eeal{ML45}
where
\begin{eqnarray}
\label{ML44x}
 &\mychithree
  :=
 \Bigg[-
  \frac
 \partial {\partial x} \left( \frac {F(x)}{(x-z)^2}\right)
 + \frac {F^2(x)} {G(x)}\frac
 \partial {\partial x} \left(
 \frac{G(x)}{F(x)(x-z)^2}\right)\Bigg]
 \frac{A^2(x-z)^4 }{G(x)F^2(z)}c_\varphi ^2 & 
\\
 &
    \phantom{=}
 -\frac{F'(x)F(z) }{ F^2(x)}c_t^2
 +
  \frac
 \partial {\partial x}\left( \frac {F(x)}{(x-z)^2}\right)
  \frac{(x-z)^{2}}{ F(x)}\Bigg(\lambda +
 \frac{ F(z)}{F(x)}c_t^{2} \Bigg)
 \;. &
 \nonumber
\end{eqnarray}
Equivalently,
\bea
 {\frac 2{\sqrt {F(x)G(x) }}
 \frac{d}{ds}\left(\frac{F^2(z) \sqrt{F(x)}}{A^2 (x-z)^2  }\frac
 {d\myrho }{ds}\right) = \mychithree}
 \;.
\eeal{xeq4}
Multiplying by $ds/dz$,  taking into account that $\mathring a
\ne 0$ by \eq{ML48} and \eq{ML49}, from  \eq{ML44} we are led
to an evolution equation of the form
\bel{evolveqx}
 \frac{d}{dz}\left(\frac{F^2(z) \sqrt{F(x)}}{A^2
(x-z)^2 }\frac {d\myrho }{ds}\right) = O(|z-\xi_3|^{-3/4})
 \;.
\ee
We note that the right-hand side of \eq{evolveqx} is integrable in
$z$ near $\xi_3$, hence $|\dot \theta|$ is bounded near $\xi_3$.

Using arguments already given, it is straightforward to obtain now
the following:

\begin{Theorem}
 \label{Tbelu}
{Causal geodesics maximally extended in the region $\{\xi_F< z
< \xi_3\}$ and directed towards $z=\xi_F$ reach this last set
in finite affine time.}
%
\end{Theorem}

In fact the conclusion holds true as well for those spacelike
geodesics for which $\dot z$ does not change sign.

To get uniformity towards $\xi_3$, we first note that, from the
integrability of the right-hand-side of \eq{evolveqx},
\bel{u2}
 \lim_{z\to\xi_3}\dot \theta \ \mbox{ exists, in
 particular} \
 |\dot \theta| \le C_\epsilon \ \mbox{ on   $[\xi_F+\epsilon,
 \xi_3]$.}
\ee

We return now to \eq{dv1}-\eq{dw1}. From what has been said,
the limit $\lim_{z\to \xi_3}\beta $ exists. By \eq{lambda5}
multiplied by $|G(z)|$, or otherwise, we find that the limit
$\lim_{z\to \xi_3}\alpha $ exists, and
\bel{albesi}
 \lim_{z\to \xi_3}\alpha  = \pm \lim_{z\to \xi_3}\beta
  \;.
\ee
We have $\lim_{z\to \xi_3}\beta \ne 0$ by \eq{ML48}. Suppose,
first, that \eq{albesi} holds with the plus sign.
We write $(\alpha-\beta)/G$ as
$[(\alpha^2-\beta^2)/G]/(\alpha+\beta)$, and  use \eq{lambda5}
to obtain that the limit
$$
 \lim_{z\to \xi_3}\frac{\alpha-\beta}{G(z)}
$$
exists. This implies that \eq{dw1} can be written in the form
\bel{ML50}
 \mydot {\hat w} = \phi(s) \hat w
\ee
for some continuous function $\phi$. By integration $\hat w$
has a non-zero limit as $z\to \xi_3$, and \eq{ML50} implies
that $\mydot {\hat w}$ has a limit as well. It follows that
$\hat v$ tends to be zero, since $\hat v \hat w$ does. \Eq{dv1}
shows now that $\mydot {\hat v}$ has a limit. Hence
\bel{firstposx2}
   \hat v + \hat w +  \left| \mydot {\hat w}\right|
  +  \left| \mydot {\hat v}\right| + \left| \mydot z \right| + \left| \mydot \myrho \right|  \le C
 \;.
\ee
%
>From what has been said so far we conclude that those geodesics
smoothly extend across the Killing horizon $\{\hat w \hat v
=0\}$.

A similar argument, with $\hat v$ interchanged with $\hat w$,
applies when the minus sign occurs in \eq{albesi}.

Summarising, we have proved:

\begin{Proposition}
 \label{Preachbif}
{Causal geodesics in the region $\{\xi_F<z< \xi_3\}$ reach the
bifurcate Killing horizon $\{\hat w \hat v =0\}$ in finite
affine time, and are smoothly extendible there.}
%
\end{Proposition}

\medskip

\subsection{Geodesics in the region $\{\xi_3<z\le 2 \xi_3\}$}
 \label{ss3t3}

In this coordinate range    $F(z)$ is strictly negative  and
$G(z)$ is positive, so we rewrite \eq{lambda} as
\begin{eqnarray}
\label{lambda2xxl}\lefteqn{ \phantom{xxxxx}
  \frac{|F(z)|}{ G(z) } {\dot{z }^{2} }
+\frac{F^2(z)}{F(x)} {\dmyrho ^{2}}  +
\frac{A^4(x-z)^4 }{G(x)F^2(z)}c_\varphi ^2
 }
 &&
 \\
&&
 ={\frac{A^4(x-z)^4}{F^2(x)  G(z)}\left(\sigma(\xi_1-z)c_t-c_\psi\right)^{2}}
 -\frac{A^{2}(x-z)^{2}}{F(x)}\Bigg[-\lambda
+\frac{| F(z)|}{F(x)}c_t^{2} \Bigg] \; .
\nonumber
\end{eqnarray}
>From this one immediately obtains a uniform bound on $\dot z$,
as well as
\begin{eqnarray}
 {
 \left|\frac {d\myrho} {ds}\right| +
\frac{c_\varphi ^2 }{ \sqrt{G(x)}} \le
  \frac{C}{\sqrt{|G(z)|}}
\;.  }
\label{15X.1}
\end{eqnarray}

Geodesics on which $z$ stays bounded away from $\xi_3$ have
already been taken care of in Section~\ref{sGdoc}.
So we consider a geodesic segment $\gamma:[0,\mathring
s)\to\{\xi_3<z\le 2\xi_3\}$, with $\mathring s < \infty$, such
that there exists a sequence $s_i\to \mathring s$ with
$z(s_i)\to \xi_3$. (If $\mathring s=\infty$ there is nothing to
prove.) But we have just seen that the function $s\mapsto z(s)$
is uniformly Lipschitz, hence $z(s)\to \xi_3$ as $s\to
\mathring s$.

The Killing vector field
\bel{KilHorpropx}  X:= \frac{\partial}{\partial t} +
\frac{A\sqrt{\xi_F }}{\sqrt{\nu}
 (\oldxifour  -
 \oldxitwo )} \frac{\partial}{\partial \psi}
\ee
is tangent to the generators of the Killing horizon $\mcE$,
thus light-like at $\mcE$. As  the horizon is non-degenerate,
$X$ is timelike near $\mcE$ for small negative values of $\hat
v \hat w$. But then $g(X,\dot \gamma)<0$ for causal future
directed geodesics in the domain of outer communications near
$\mcE$, which shows that for causal geodesics through the
region of current interest we must have
\bel{restcondnot}
 g(X,\dot \gamma) = c_t + \frac{A\sqrt{\xi_F
 }}{\sqrt{\nu}
 (\oldxifour  -
 \oldxitwo )}c_\psi \ne 0
 \quad
\Longleftrightarrow \quad
 \sigma(\xi_1-\xi_3)c_t-c_\psi\ne 0
 \;.
\ee
It follows that causal geodesics intersecting $\doc$ for which
\eq{restcondnot} does \emph{not} hold stay away from a
neighborhood of $\mcE$, and are therefore complete by the
results in Section~\ref{sGdoc}.

>From what has been said we conclude that:

\begin{Proposition}
 \label{Pcrosscond}
{Causal geodesics crossing the event horizon $\{\hat v \hat
w=0\}$ and satisfying
$$
 \sigma(\xi_1-\xi_3)c_t-c_\psi=0
$$
are entirely contained within $\{\hat v \hat w>0\}\cup \{\hat
v=0= \hat w\} $.}
%
\end{Proposition}

We continue by noting that \eq{zeq2ag4} remains true with
$\mychione$ as given by \eq{MLIx1} on every interval of values
of $s$  on which $\dot z$ does not change sign.
We want to show, using a contradiction argument, that $\dot z$
will be eventually negative for $s$ close enough to $\mathring
s$. So suppose not, then there will be increasing sequences
$\{s_i^\pm\}_{i\in \N}$, $s_i^\pm \to \mathring s$, with
$s_i^-<s_i^+$, such that
$$
  \dot z(s_i^\pm)=0\;, \  \dot z(s)<0 \ \mbox{on} \ I_i:=(s_i^-,s_i^+)
\;, \  z_i^\pm :=z(s_i^\pm)  \searrow \xi_3\;, \ z_i^->z_i^+
 \;.
$$
By inspection of \eq{MLIx1} and \eq{zeq2ag5}, there exists
$z_*>\xi_3$ and $\epsilon>0$ such that for all $z\in
(\xi_3,z_*)$ we have
\bel{15X.5}
  \frac {dh^2}{dz} \le  -\frac{\epsilon}{ (z-\xi_3)^2}
 \;.
\ee
Integrating,
\bel{15X.6}
  h^2 (z_i^-) -  h^2 (z_i^+) \le  -\int_{z_i^+}^{z_i^-} \frac{\epsilon}{
(z-\xi_3)^2} dz <0
 \;,
\ee
which contradicts the fact that $ \dot z(s_i^\pm)=0$, and shows
that $\dot z$ is indeed strictly negative sufficiently close to
the event horizon.

As in Section~\ref{ssUb} we obtain now \eq{ML41a}. This,
together with the definition of $h$, implies that $dz/ds$ is
strictly bounded away from zero; equivalently,
$$
 \left| \frac{ds}{dz}\right | \le C
 \;.
$$
In particular $\{\hat v \hat w = 0 \}$ is reached in finite
affine time. Moreover, from \eq{15X.1} we now find
\begin{eqnarray}
 {
 \left|\frac {d\myrho} {dz}\right|   \le
  \frac{C}{\sqrt{|G(z)|}}
\;.  }
\label{15X.1x}
\end{eqnarray}
Keeping in mind \eq{restcondnot}, one can now repeat the
arguments of Section~\ref{ssUb} (after \eq{ML48}, with
\eq{lambda2xxxx} replaced by \eq{15X.1x}),
to conclude that:

\begin{Theorem}
 \label{Tgeoddoc}
{All maximally extended causal geodesics through $\doc$ are
either complete, or can be smoothly extended across the horizon
$\{\hat v \hat w =0 \}$.}
%
\end{Theorem}

\medskip

\subsection{The Killing horizon}

Consider a causal geodesic $\gamma$ such that $\gamma(s_0)
\in\{\hat v \hat w =0 \}$. If $d \hat v /ds$ or  $d \hat w /ds$
are both different from zero at $s_0$, then $\gamma$
immediately enters the regions already covered. If $\gamma$
enters $\{z < \xi_3\}$ it will stay there by monotonicity of
$z$, so Theorem~\ref{Tbelu} applies.
Otherwise it enters the region $\{z > \xi_3\}$; then either it
approaches $z=\xi_3$ again, in which case it crosses back to
$\{z < \xi_3\}$ by Theorem~\ref{Tgeoddoc}, and
Theorem~\ref{Tbelu} applies; or it stays away from $z=\xi_3$
for all subsequent times, in which case it is complete, again
by Theorem~\ref{Tgeoddoc}.

So it remains to consider geodesics for which
$$
 \forall \ s \quad  {\hat v}(s)  {\hat w}(s) =0 = \mydot {\hat v}(s) \mydot {\hat w}(s)
 \;.
$$
Since the bifurcation ring $S:=\{\hat w = \hat v = 0\}$ is
spacelike, those causal geodesics which pass through $S$
immediately leave the bifurcate horizon, except for the
generators of the latter. But those generators are complete by
standard results~\cite{Boyer}. Since the only causal curves on
a null hypersurface are its generators, the analysis is
complete.

This achieves the proof of Theorem~\ref{Tglobcontge}.
\qed

\bigskip

\bibliographystyle{amsplain}\def\cprime{$'$} \def\cprime{$'$} \def\cprime{$'$} \def\cprime{$'$}
\providecommand{\bysame}{\leavevmode\hbox
to3em{\hrulefill}\thinspace}
\providecommand{\MR}{\relax\ifhmode\unskip\space\fi MR }
\providecommand{\MRhref}[2]{%
  \href{http://www.ams.org/mathscinet-getitem?mr=#1}{#2}
} \providecommand{\href}[2]{#2}

\end{document}